\documentclass[12pt,preprint]{emulateapj}
\usepackage{epsfig}

\newcommand{\kms}{\mbox{km s$^{-1}$}}

\newcommand{\intfluxmks}{\mbox{Wm$^{-2}$}}

\newcommand{\ha}{\mbox{H$\alpha$}}
\newcommand{\hb}{\mbox{H$\beta$}}

\newcommand{\lam}{\mbox{$\lambda$}}

\newcommand{\msun}{\mbox{M$_{\odot}$}}
\newcommand{\Lsun}{\mbox{L$_{\odot}$}}

\shorttitle{Detailed portrait of a $z\sim 2.6$ SMG}

\begin{document}

\title{Intense Star-formation and Feedback at High Redshift:
Spatially-resolved Properties of the ${\rm z}=2.6$ Submillimeter Galaxy
SMMJ14011$+$0252\altaffilmark{1}}

\author{N.~P.~H.~Nesvadba\altaffilmark{2,3}, 
M.~D.~Lehnert\altaffilmark{2,3},
R.~Genzel\altaffilmark{2,4}, 
F.~Eisenhauer\altaffilmark{2}, 
A.~J.~Baker\altaffilmark{5,6},
S.~Seitz\altaffilmark{2,7},
R.~Davies\altaffilmark{2}, 
D.~Lutz\altaffilmark{2},
L.~Tacconi\altaffilmark{2}, 
M.~Tecza\altaffilmark{8},
R.~Bender\altaffilmark{2,7}, 
R.~Abuter\altaffilmark{9}}

\altaffiltext{1}{Based on observations collected at the European
Southern Observatory, Very Large Telescope Array, Cerro Paranal, Chile
(70.A-0254(A), 70.B-0545(A), and 073.A-5028(A)).}
\altaffiltext{2}{Max-Planck-Institut f\"ur extraterrestrische Physik,
Giessenbachstra\ss e, D-85748 Garching bei M\"unchen, Germany.}
\altaffiltext{3}{Current Address: Laboratoire Galaxies, Etoiles,
  Physique et Instrumentation, Observatoire de Paris, 5 place Jules
  Janssen, 92195 Meudon, France}
\altaffiltext{4}{Department of Physics, University of California,
Berkeley, CA 94720.}
\altaffiltext{5}{Jansky Fellow, National Radio Astronomy Observatory,
Department of Astronomy, University of Maryland, College Park, MD
20742-2421.}
\altaffiltext{6}{Current Address: Department of Physics and Astronomy
Rutgers, the State University of New Jersey 136 Frelinghuysen Road
Piscataway, NJ 08854-8019}
\altaffiltext{7}{Universit\"{a}ts-Sternwarte, Scheinerstrasse 1, M\"unchen
D-81679, Germany.}
\altaffiltext{8}{University of Oxford, Subdepartment of Astrophysics,
Denys Wilkinson Building, Keble Road, Oxfordshire, Oxford OX1 3RH, UK}
\altaffiltext{9}{European Southern Observatory, Karl Schwarzschild
Stra\ss e, D-85748 Garching bei M\"unchen, Germany}

\begin{abstract} 
We present a detailed analysis of the spatially-resolved properties of
the lensed submillimeter galaxy SMMJ14011$+$0252 at z=2.56, combining
deep near-infrared integral-field data obtained with SPIFFI on the VLT
with other multi-wavelength data sets.  As previously discussed by
other authors, the broad characteristics of SMMJ14011$+$0252 in
particular and submillimeter galaxies in general are in agreement with
what is expected for the early evolution of local massive spheroidal
galaxies. From continuum and line flux, velocity, and dispersion maps,
we measure the kinematics, star-formation rates, gas densities, and
extinction for individual subcomponents. The star formation intensity
is similar to low-redshift ``maximal starbursts'', while the line
fluxes and the dynamics of the emission line gas provide direct
evidence for a starburst-driven wind with physical properties very
similar to local superwinds.  We also find circumstantial evidence for
"self-regulated" star formation within J1. The relative velocity of
the bluer companion J2 yields a dynamical mass estimate for J1 within
$\sim 20$ kpc, $M_{dyn} \sim 1 \times 10^{11}$ \msun.  The relative
metallicity of J2 is 0.4 dex lower than in J1n/s, suggesting different
star formation histories.  SED fitting of the continuum peak J1c
confirms and substantiates previous suggestions that this component is
a z$=$0.25 interloper. When removing J1c, the stellar continuum and
\ha\ line emission appear well aligned spatially in two individual
components J1n and J1s, and coincide with two kinematically distinct
regions in the velocity map, which might well indicate a merging
system. This highlights the close similarity between SMGs and ULIRGs,
which are often merger-driven maximal starbursts, and suggests that
the intrinsic mechanisms of star-formation and related feedback are in
fact similar to low-redshift strongly star-forming systems.
\end{abstract}


\section{Introduction}
\label{sec:introduction}

Half of the stars in the local universe formed at $z\gtrsim 1$ or
by about half the age of the universe \citep[e.g.,][and references
therein]{rudnick03}.  However, in spite of our knowledge of ``when'',
our understanding of the physical processes triggering and governing this
star formation is still rudimentary. In particular, does the physics
of star formation depend on redshift? For example, \citet{goldader02}
hypothesize that the intense star formation observed in some high-redshift
galaxies might require a more efficient mode to form stars, without
triggering by major mergers. \citet{erb06}, on the other hand,
find that star formation in galaxies at z$\sim$2 is less efficient,
because they drive very efficient winds and energy into galaxy halos
and the inter-galactic medium (IGM) and thus lose significant amounts
of mass.  Both hypotheses illustrate the need for detailed studies of
star formation at high-redshift, to investigate whether low-redshift
star-forming galaxies observed at high spectral and physical resolution
can be good analogs to star formation at high redshift.

``Superwinds'' -- vigorous outflows of hot gas due to the thermalized
ejecta of supernovae and massive stars in starburst galaxies -- are
inexorably linked to star formation. While winds likely play a fundamental
role in galaxy formation and evolution \citep[e.g.,][]{heckman90,
lehnert96a}, and contribute significantly to the metal content of
the IGM \citep[e.g.,][]{bouche05}, the direct observational evidence
for the ubiquity of winds in star-forming galaxies at high redshift
is still rather limited. It is mainly based on blue line asymmetries
and offsets of rest-frame UV interstellar absorption lines relative to
optical emission lines \citep[e.g.,][]{pettini01} and on the evolution
of the mass-metallicity relationship at high redshift \citep{erb06}
in UV-selected galaxy populations.

Submillimeter galaxies (SMGs) at z$\sim$ 2 are the sites of particularly
vigorous star formation, with star-formation rates of $\sim 100-1000$
\msun\ yr$^{-1}$, and thus provide an excellent opportunity to investigate
the properties of extreme star formation at high redshift \citep[][
and references therein]{blain02}.

The z$=$2.57\footnote{Using the flat $\Omega_{\Lambda} =0.7$ cosmology
with H$_0=70$ kms$^{-1}$ Mpc$^{-1}$ leads to D$_L=21.04$ Gpc and
D$_A=1.66$ Gpc at $z=2.565$. The size scale is 8.03~kpc/\arcsec. The age
of the universe for this redshift and cosmological model is 2.5 Gyrs.}
SMG SMMJ14011+0252 \citep{ivison01, frayer99} is perhaps the best
studied SMG across all wavebands, because it has a relatively bright
multiwavelength contiuum, rest-frame optical emission lines with rather
large equivalent widths, and moreover, is moderately gravitationally
lensed (${\cal M}$$\lesssim$5) by the z$=$0.25 galaxy cluster A1835
\citep[see][]{ivison01, barger99, fabian00, ivison00, smail00, smail02,
downes03, tecza04, swinbank04, motohara05, smail05}. \citet{ivison00} gave
the first detailed description of the source properties, and estimate
a far-infrared luminosity of ${\cal L}_{FIR} \sim 6\times 10^{12}$
L$_{\odot}$ and a star-formation rate SFR$=$1260-3900 M$_{\sun}$ yr$^{-1}$
\citep[for a magnification of 3; see also][]{ivison01}. \citet{tecza04}
re-analyzed the FIR data given in \citet{ivison00}, estimating ${\cal
L}_{FIR}\sim 2.3 \times 10^{13} {\cal M}^{-1}$ L$_{\odot}$ and FIR
star-formation rates $\sim 1920 {\cal M}^{-1}$ \msun\ yr$^{-1}$, where
${\cal M}$ indicates the magnification by the gravitational lens. The
differences are mainly due to different assumptions regarding the
modelling of the spectral energy distributions in the far-infrared and
different initial mass functions for the star-formation rates. Since the
values used in \citet{tecza04} are better matched to the assumptions made
in our paper, we will in the following use the values of \citet{tecza04}.

Many SMGs show evidence for optically evident AGN through their rest-frame
optical line ratios \citep{takata06}. \citet{swinbank05} interpret the
broad recombination line profiles in some SMGs as likely originating from
nuclear broad line regions. Deep photometry of SMMJ14011+0252 covers
X-ray to radio wavelengths, and longslit spectroscopy has previously
been taken in the rest-frame UV and optical range. None of these data
have revealed evidence for an AGN in this source. All these arguments
make SMMJ14011+0252 an ideal target to study the properties of strongly
star-forming galaxies in the early universe.

The relationship of the molecular gas relative to other
components of SMMJ14011+0252 however has led to controversy in the
literature. \citet{frayer99} reported the first detection of CO(3-2)
emission, and \citet{ivison01} emphasize a good alignment of
the CO(3-2) emission with the faint red component J1n, indicating
that this is the location of the intense submm emission and the
starburst (see Fig.~\ref{fig:CO} for the labeling). This view was
later challenged by \citet{downes03}, who found a different alignment
of their CO(7-6) and CO(3-2) data sets with the rest-frame UV data,
placing the CO emission significantly outside the optical emission of
J1. The CO emission is marginally resolved spatially \citep[2\arcsec\
$\times$$<$0.4\arcsec;][]{downes03}, and its line width of $FWHM=190\pm
10$ \kms\ corresponds to a dynamical mass of $3\times 10^9$ \msun\
\citep[][not correcting for inclination]{greve05}.  We note that this
mass estimate is less than the estimate of the total molecular mass of
SMMJ14011$+$0252 \citep{frayer99, downes03}.

High-resolution HST F702W imaging shows that SMMJ14011$+$0252 has a
complex morphology in the rest-frame UV, which does not become more
regular in rest-frame optical wavebands \citep{ivison00}. Based on
rest-frame UV spectroscopy, \citet{ivison00} were the first to argue
that the nearby blue component J2 is at a very similar but not identical
redshift, and is a physically related component of J1. The nature of
the bright UV-optical continuum peak J1c remained more mysterious:
\citet{downes03} proposed that J1c might be a member galaxy of the
foreground cluster A1835 along the line of sight and suggested that
several blue features of J1 were in fact multiple images caused by strong
lensing through a foreground source J1c. \citet{smail05, swinbank04}
agree with the interpretation that J1c is a foreground source, but
favor a lower magnification factors ${\cal M}\sim 3-5$, which would
not lead to multiple images or strong differential lensing. They argue
that absorption lines seen in the spectrum of \citet{barger99} can be
identified as rest-frame optical absorption lines at z$\sim$0.25, and
highlight the nearly spherical morphology of J1c, both of which seem
unlikely, if it is a lensed galaxy at z$=$2.57.

In a previous paper \citep[][hereafter Paper~I]{tecza04}, we presented
an initial analysis of deep near-infrared integral field spectroscopy
of SMMJ14011+0252 obtained with SPIFFI on the VLT, concentrating on the
integrated properties of the source, such as the high gas-phase oxygen
abundance (12$+$[O/H]$=$9.0, measured with $R_{23}$) and large mass, and
put these results into a broader perspective of mass assembly in the early
universe within hierarchical structure formation models. We complement
these findings now through a detailed analysis of the spatially-resolved
properties of this source, and we particularly discuss the implications
for star formation and related feedback in massive galaxies at high
redshift.

\section{SPIFFI and Complementary Data Sets} \label{sec:obsdatred}

SMMJ14011+0252 was observed in the J, H and K-bands with the integral
field spectrometer SPIFFI \citep{eisenhauer03} at the ESO VLT in
spring 2003, with individual exposure times of 300 s in H, and 600 s in J
and K. Total exposure times were 340 min in K, 95 min in H and 60 min in
J. We used the 0.25\arcsec\ pixel scale and obtained spectral resolutions
of R=$\frac{\lam}{\Delta\lam}\sim$1500 in the J-, $\sim$2000 in the H-,
and $\sim$2400 in K-bands.

The data were calibrated using the UKIRT faint standard FS135 and
agree to within a few percent with the magnitudes previously published
for SMMJ14011$+$0252 \citep{ivison00}. The seeing disk has
FWHM$=0.6$\arcsec $\times$ 0.4\arcsec\ in right ascension and
declination, respectively and was measured from the standard star. The
data reduction has been described elsewhere \citep[][]{nesvadba06a}.
The main difference between the data reduction in this paper and
Paper~I is improved sky subtraction allowing us to investigate the
spatially-resolved properties of SMMJ14011$+$0252 more robustly. The
previous reduction algorithm led to an effective oversubtraction of
the sky background near bright emission lines. As a result, the data
quality in individual frames was improved through the new reduction,
which yielded a more accurate alignment of the individual frames,
which led to a better data quality in the combined data set and a
better spatial resolution. This has no significant impact on the
integrated spectrum discussed in \citet{tecza04}, but does change the
mapping of kinematics and emission line morphologies by improving the
signal-to-noise of the low surface brightness line emitting regions.

In addition, before and since the publication of Paper I, a large
number of complementary data sets are (now) available to supplement and
refine the interpretation of the SPIFFI data and the overall nature
of the source. \citet{ivison01} obtained high-resolution optical
imaging of the A1835 field through the F702W ($\equiv R_{702}$)
filter of the WFPC2 on-board the HST. The reduced image was kindly
provided by R.~Ivison. He also kindly shared his 1.4\,GHz continuum
map of A1835 with us, which was obtained with the NRAO Very Large Array
\citep[VLA;][]{ivison01}. D. Downes kindly provided us with his CO (3--2)
and 242\,GHz continuum maps of SMMJ14011$+$0252 obtained with the IRAM
Plateau de Bure Interferometer \citep[PdBI; ][]{downes03}.

Deep ISAAC J, H, and K band images are available from the ESO archive and
were reduced as described in \citet{bremer04}. A deep VLT FORS1 V-band
image of the field of A1835 was obtained, reduced, and is described
in \citet{lehnert05}. A. Barger kindly provided the rest-frame UV
spectra taken with LRIS on the Keck 10m telescope and published in
\citet{barger99}. Finally, R. Pello and A. Hempel gratiously shared their
F850LP ($\equiv Z_{850}$) ACS image of SMMJ14011$+$0252 J1 and J2 before
the data were publicly available.

\section{Absolute Astrometry}
\label{sec:astrometry}

Various possible and disparate alignments have been proposed for
SMMJ14011$+$0252 in the literature \citep[cf.][]{downes03,ivison01}.
The position of both the radio and mm interferometric positions relative
to the rest-frame optical and UV have been especially problematic.

\begin{figure}
\epsscale{0.7}
\plotone{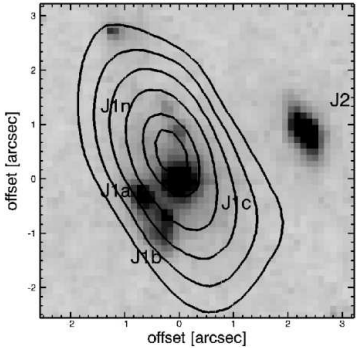}
\caption{Contours showing the distribution of the CO(3-2) line emission
superimposed on the F702W WFPC2 broad band continuum image (shown in
grayscale).  The CO(3-2) map is from \cite{downes06} and the labels on the
WFPC2 image designate the regions of SMM14011$+$0252 following the scheme
of \citet{motohara05}.  J1n refers to the diffuse extended region north
of J1c (see also Fig.~\ref{fig:CO_jk}). North is up and east is to the
left. The astrometry used for this overlay was derived from identifying
6 radio sources in the FORS V-band image (see text for details).
\label{fig:CO}}
\end{figure}

\begin{figure}[hbt] 
\epsscale{0.7}
\plotone{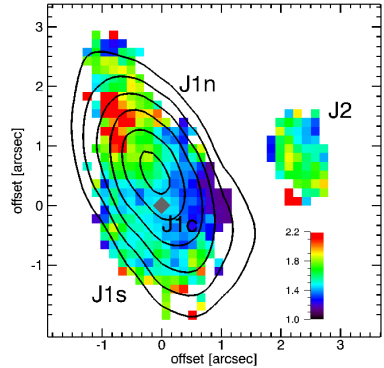}
\caption{The J$-$K color distribution with superimposed contours
showing the CO(3-2) flux distribution map. North is up and east is to
the left. From 
our own lensing models and those of \citet{smail05}, the direction of
the largest lensing shear is roughly along the axes of largest extent
in J1 and J2. J band emission in J1n is below the 3$\sigma$ limit in
$\sim$10\% of the pixels where K band emission is detected. The
reddest area in J1n therefore represents a 3$\sigma$ lower limit on
the intrinsic J$-$K color.
\label{fig:CO_jk}}
\end{figure}

\begin{figure}
\epsscale{0.7}
\plotone{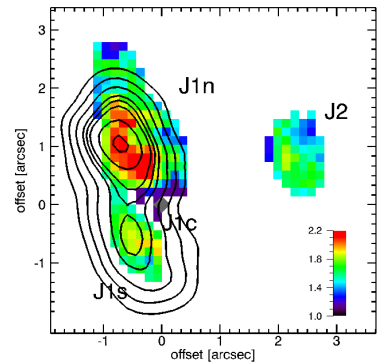}
\caption{J$-$K color distribution of SMMJ14011$+$0252 as in
Fig.~\ref{fig:CO_jk} but with the model of J1c removed. Contours
indicate \ha\ line emission in steps of 2$\sigma$ with the lowest contour
representing 3$\sigma$.  The contribution of strong H$\alpha$ emission
has been removed when calculating the J$-$K colors. North is up, east
to the left.
\label{fig:jmink}}  
\end{figure}

We obtained a new and more robust astrometry for the field of
SMMJ14011$+$0252 using ISAAC K-band and FORS V-band imaging of A1835
\citep{lehnert05}. We identified six radio sources in the V-band image
at magnitudes brighter than 17.5 mag (the V-band image has a 5$\sigma$
limit of 27.8 mag).  Of these 6 sources, 3 were point sources at the
resolution of the data (seeing FWHM$\sim$0.7-0.8 arcsec) and 3 were
extended.  Given their high signal-to-noise, the uncertainty in the
position is very small ($\la$0.08 arcsec). The depth and field of view
of this image are greater than previously available
optical/near-infrared data.  The radio sources are spread over the
roughly 6$\arcmin\times$6$\arcmin$ field of view of the V-band image.
A comparison of the radio source and optical positions suggests that
the best alignment has an RMS scatter of $0.04\arcsec \times
0.08\arcsec$ ($\sim 10\times$\ less than the absolute positional
uncertainty in each band).  Such a good alignment suggests that both
the position in each frame can be accurately determined and that the
relative distortion in both the radio map and V-band image are
insignificant compared to the total relative positional uncertainties
over the entire field of view of the V-band image.  All other
optical/NIR data can be referenced to the V-band image to a very high
accuracy given the large number of sources that we used for the
relative alignment (several dozen point or point-like sources).  The
VLA radio and PdBI millimeter maps share the ``radio'' coordinate
frame. We estimate a total absolute uncertainty of $0.33\arcsec \times
0.34\arcsec$\ ($\lesssim 1.5$ SPIFFI pixels) in right ascension and
declination of the optical and near-infrared data relative to the CO
position, including an absolute astrometric uncertainty in the CO map
of 0.3\arcsec\ \citep[following][]{downes03} and a fiducial
uncertainty of 0.14\arcsec\ (the pixel size of the ISAAC data, which
is certainly an over-estimate of the true uncertainty since the
relative alignment is better than a pixel).  The relative uncertainty
between the VLA map and the optical/near-infrared data is smaller (by
a factor of 4 to 8).

The aligned K-band ISAAC image was used to put the SPIFFI cubes into
the common frame, based on the positions of J1 and J2 and assuming a
pixel scale of 0.25\arcsec\ for the SPIFFI data. J1 and J2 have
signal-to-noise ratios $>$15 and 6 in the SPIFFI continuum image in
the central pixel, respectively, and signal-to-noise ratios $>$30 and
11 in the ISAAC K band image, respectively.  The alignment is limited
by the uncertainty of the peak position in J1c and J2, which is much
smaller than a single pixel in either data set (0.25\arcsec\ for
SPIFFI or 0.14\arcsec\ for ISAAC).

Our new astrometric alignment rules out previous claims that the
CO emission might be significantly offset from the UV and optical
positions, and is an independent confirmation of the inital alignment
of \citet{ivison01}, based on some new data sets. The position of the
CO data is shown in Fig.~\ref{fig:CO} as contours overlaid on the HST
$R_{702}$ image, and in Fig.~\ref{fig:CO_jk} with respect to the ISAAC
J$-$K color image.  Within the uncertainty of $0.33\arcsec$ the CO emitter
can be identified with either J1n or J1c. The redder colors and stronger
star formation (\S\ref{subsec:sfintens}) favor J1n as the more likely
source of CO line emission. We summarize the positions of the
individual components in SMMJ14011$+$0252 relative to the radio frame
in Table~\ref{tab:astrometry}.  At any rate, this calibration firmly
places the CO and radio emission within the isophotes of J1 in the HST
\citep{ivison00} and ground-based images \citep{ivison00, bremer04,
lehnert05}.

\section{Results and Analysis}
\label{sec:results}

\subsection{Continuum Morphology and Colors}
\label{subsec:continuum}

The rest-frame optical and UV continuum and \ha\ emission line
morphology of the J1 complex have been described elsewhere \citep[see,
e.g.][]{ivison00,tecza04,motohara05,smail05}.  Overall, our data are
consistent with these previous descriptions, but deeper and more
detailed in several aspects. We label individual components in
Fig.~\ref{fig:CO}$-$\ref{fig:jmink}.

\begin{deluxetable}{lccccc}
  \tablecolumns{8}
  \tablecaption{SMMJ14011$+$0252: Astrometry}
  \tablehead{\colhead{Component} & \colhead{R.A.(2000)} & \colhead{Decl.(2000)}
    & \colhead{$\Delta^{J1c}_{RA}$} & \colhead{$\Delta^{J1c}_{Dec}$} &
  \colhead{Data} \\ 
    & (2) & (3) & (4) & (5) & (6) }
\tablecomments{Absolute positions in the radio frame of subcomponents
  from our revised astrometry. Uncertainties in the relative
  alignement of the CO are $\sim 0.33\arcsec$.  Column (1) --
  Designations of the various regions defined with SMMJ14011$+$0252.
  In the text, since a majority of the data sets do not resolve J1a
  and J1b, we have used the designation J1s for the combined and
  surrouding emission of J1a and J1b. Column (2) -- Right
  ascension. Column (3) -- Declination. Column (4) -- Relative offset
  to J1c in right ascension in arcsec. Column (5) -- Relative offset
  to J1c in declination in arcsec. Column (6) -- Data set used for
  measurement. F702W -- WFPC2 data of \citet{ivison00}. I01 -- VLA 1.4
  GHz radio data of \citet{ivison01} -- D06 -- PdBI CO data of
  \citet[][and private communication of currently unpublished data
  sets]{downes06}. SPIFFI -- K-band IFU data set.}

\startdata
  J1c       & 14:01:04.933 & 02:52:23.98 & 0.0  & 0.0   & F702W \\
  J1a       & 14:01:04.967 & 02:52:23.73 & 0.51 & 0.25  & F702W \\
  J1b       & 14:01:04.948 & 02:52:23.43 & 0.22 & 0.55  & F702W \\
  J1n       & 14:01:04.969 & 02:52:24.81 & 0.54 & 0.83  & SPIFFI \\
  1.4GHz    & 14:01:04.92 & 02:52:24.80  & 0.20 & 0.83  & I01  \\
  PdBI 1mm  & 14:01:04.933 & 02:52:24.20 & 0.00 & 0.22  & D06 \\
  PdBI 3mm  & 14:01:04.933 & 02:52:24.37 & 0.00 & 0.39  & D06 \\
  \hline
  J2        & 14:01:04.806 & 02:52:24.73 & 1.90 & 0.75  & F702W 
\enddata
\label{tab:astrometry}
\end{deluxetable}

Continuum emission is centered on J1c and varies strongly with
wavelength. At short wavelengths J1c is small and highly symmetric. At
longer wavelengths the extended emission becomes relatively
stronger. In Fig.~\ref{fig:CO_jk} we show the J$-$K band color
distribution derived from the ISAAC data. The K band data were
convolved with a two-dimensional Gaussian disk to have the same
resolution as the J band data.  K band fluxes are corrected for the
\ha\ equivalent width measured with SPIFFI. Line contamination is
$<$0.2 mag in J1n and insignificant in J1s.

J$-$K colors (corresponding to roughly $U-R$ at rest-frame) are different
in J1n and J1s, with J$-$K$\approx 1.6$ in J1s. J band emission in J1n is
relatively faint. We replaced the J band flux by the $3\sigma$ limit where
we detected K, but no J band emission, so that the measured J$-$K$\approx
2.0$ in J1n is in fact a lower bound to the J$-$K color in J1n.

The red area in J1s coincides spatially with the bright knots J1a and
J1b in the HST F702W image. We also identify these knots in the ISAAC
K-band image. Comparison of the seeing-matched ISAAC and HST data sets
indicates that the prominence of these knots in the HST data compared
to the ground based images is in large parts due to the smaller PSF in
the F702W image and does not reflect intrinsic variations in size with
wavelength or strong line contamination. Unresolved blue knots
superimposed on J1n in Fig.~\ref{fig:jmink}, however, should also
appear at longer wavelength in spite of the lower resolution. We do
not observe them while the low surface brightness region becomes more
prominent in J1n. This indicates that these color variations reflect
intrinsic variations in the morphology of J1n.

\subsubsection{Line Ratio Diagnostics: Ionizing Source and Gas Densities}
\label{subsec:bpt}

Line emission in J1 reaches the highest surface brightnesses in the
areas of J1n and J1s, and is generally lower in J1c, as expected for a
foreground galaxy. To differentiate between J1n and J1s, we 
assign all spatial pixels to J1n and J1s, if they are at least two
SPIFFI pixels (0.5\arcsec, roughly corresponding to the seeing disk)
north or south from J1c, respectively. The spectral range of our
observations includes all strong optical emission lines between the
[OII]\lam3726,3729 and the [SII]\lam6717,6731 doublets. We use
emission line ratios, namely [OIII]/\hb, [NII]/\ha, and [SII]/\ha\ to
constrain the ionizing source \citep[e.g.][]{baldwin81}.  All line
ratios in the two-dimensional diagnostic diagrams are within the
region of HII regions for all components. J1n falls on the dividing
line between HII regions and LINERS \citep{osterbrock89}, similar to
many other high-redshift sources showing high excitation
\citep{erb06,vandokkum05}. This indicates that the bulk of the line
emission in J1n and J1s arises from gas photoionized by hot massive
stars and supports earlier X-ray measurements \citep{fabian00}, that
the far-infrared emission from this system is powered by the
starburst, without a measurable AGN component.

The \ha\ emission lines in J1n and J1s have a relative spectral shift
of $61\pm9$ \kms, significant at the level of $\sim
6\sigma$. [NII]/\ha\ ratios are $0.52\pm 0.03$ and $0.43\pm0.02$ in
the integrated spectra of J1n and J1s, respectively. Line widths (and
1$\sigma$ scatter of individual line width measurements) are FWHM$=198
\pm 32$ \kms\ in J1n, and are marginally narrower in J1s (FWHM$=157
\pm 35$ \kms) with large scatter across both components. We summarize
the emission line properties in Table~\ref{tab:emissionlines}.

\begin{figure}[hbt] 
  \centering 
  \epsfig{figure=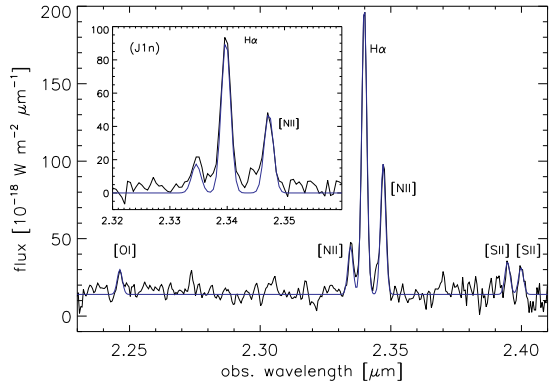,width=0.3\textwidth,angle=90}
  \caption{Integrated K band spectrum of J1n, showing the detected
    emission lines: [OI]$\lambda$6300, [NII]$\lambda$6548, \ha,
    [NII]$\lambda$6583, and the [SII]$\lambda\lambda$6717,6730
    doublet. Gaussian line fits are shown in blue, for all lines we
    assumed the FWHM and redshift measured for \ha. The inset shows a
    zoom onto the \ha\ and [NII] lines. Note the blue wings of \ha\ and
    [NII]$\lambda$6583. \label{fig:kspec}} 
\end{figure}

We calculate the extinction in the two regions from the observed \ha/\hb\
ratios (see Table~\ref{tab:emissionlines}), the Balmer decrement, and
a Galactic extinction law, as described in Paper~I, and extract the
\hb\ line emission from the same regions as \ha. For J1n and J1s, we
find $E(B-V)=1.6$ and 1.3, respectively.

\begin{figure}[htb]
  \centering
  \epsfig{figure=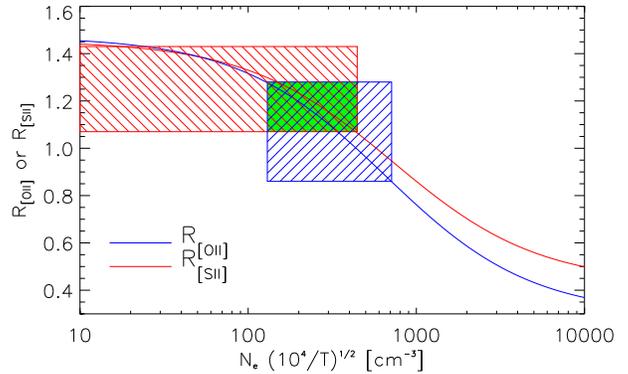,width=0.35\textwidth,angle=90}
\caption{Density estimates in SMMJ14011$+$0252 J1n as a function of
line ratio for the [OII]\lam\lam3726,3729, R$_{[OII]}$ (blue line), and
[SII]\lam\lam6716,6731, R$_{[SII]}$ (red line), doublets, scaled to a
temperature T$_e$=10$^4$ K. The red and blue shaded regions show the
ranges of line ratios and densities for the [OII] and [SII] doublets
respectively.  The green shaded region shows the overlap of best fit
density considering both doublets.  The sizes of the boxes along the
ordinate represent the range of 1$\sigma$ uncertainties in each line ratio
while the sizes of the boxes along the abscissa represent the range of
densities permitted by their respective curves of density versus R.
\label{fig:densities}} 
\end{figure}

The [OII]\lam3726,3729 and [SII]\lam6717,6731 doublets are faint, but
are spectrally resolved and have sufficient signal-to-noise ratios in
the integrated spectrum to robustly measure the ratios of the two
components, $R_{[SII]} = I(6717)/I(6731) = 1.25\pm0.18$ and
$R_{[OII]}=I(3726)/I(3729)= 1.07\pm0.21$ for [SII] and [OII],
respectively (see Table~\ref{tab:emissionlines}). For densities
between $100$ and $10^5$ cm$^{-3}$ and a given temperature, the ratio
of the two components yields the electron densities. Assuming a
``canonical'' temperature $T=10^4$ K for HII regions
\citep{osterbrock89}, our best-fit density estimate $N_e \sim 400$
cm$^{-3}$ (with favored values between $\sim$180 and 900 cm$^{-3}$,
see Fig.~\ref{fig:densities})

\subsection{Morphology and Kinematics of the Emission Line Gas: Evidence
for a Merger?} \label{subsec:gas}

Fig.~1 of Paper~I shows the continuum-subtracted \ha\ line image extracted
from the SPIFFI data cube. \ha\ emission is extended both along and
perpendicular to the direction of lensing shear in the two components
of SMM14011$+$0252, J1 and J2.  \ha\ line emission is detected in J1n
and J1s.

\begin{figure}
  \centering
  \epsfig{figure=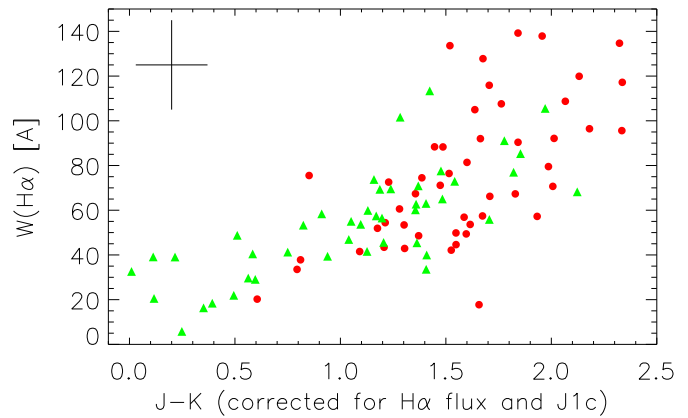,width=0.35\textwidth,angle=90}
\caption{Rest-frame equivalent widths (in \AA, positive values indicate
line emission) plotted versus the J-K color.  The contribution of strong
H$\alpha$ emission has been removed when calculating the J-K color, and
we used the J1c removed colors. Pixels that comprise J1s, and J1n, are
shown as solid green triangles, and solid red circles, respectively. The
cross indicates the typical uncertainties. \label{fig:eqwvsjmk}}
\end{figure}

\begin{figure}
\epsscale{0.7}
\plottwo{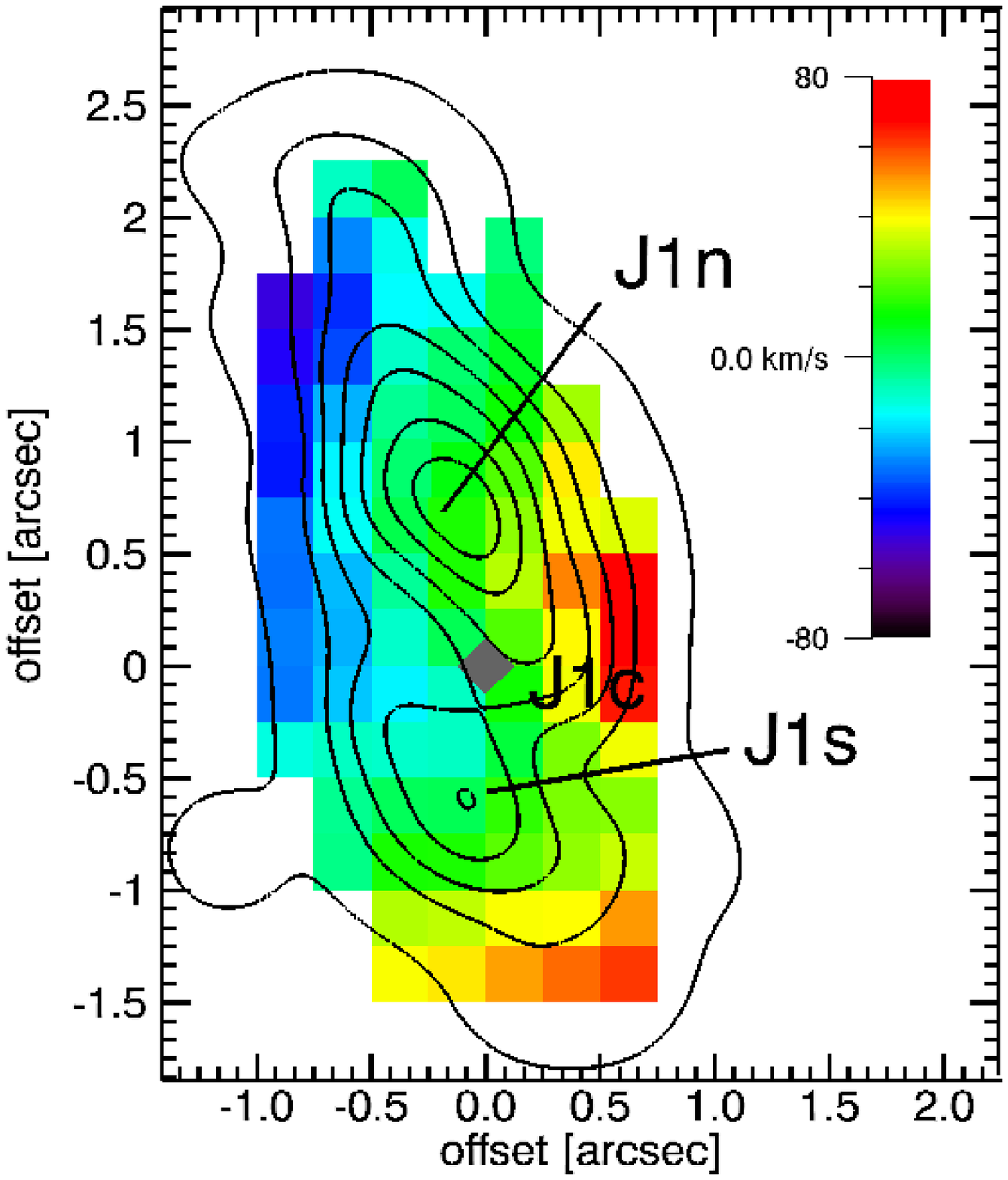}{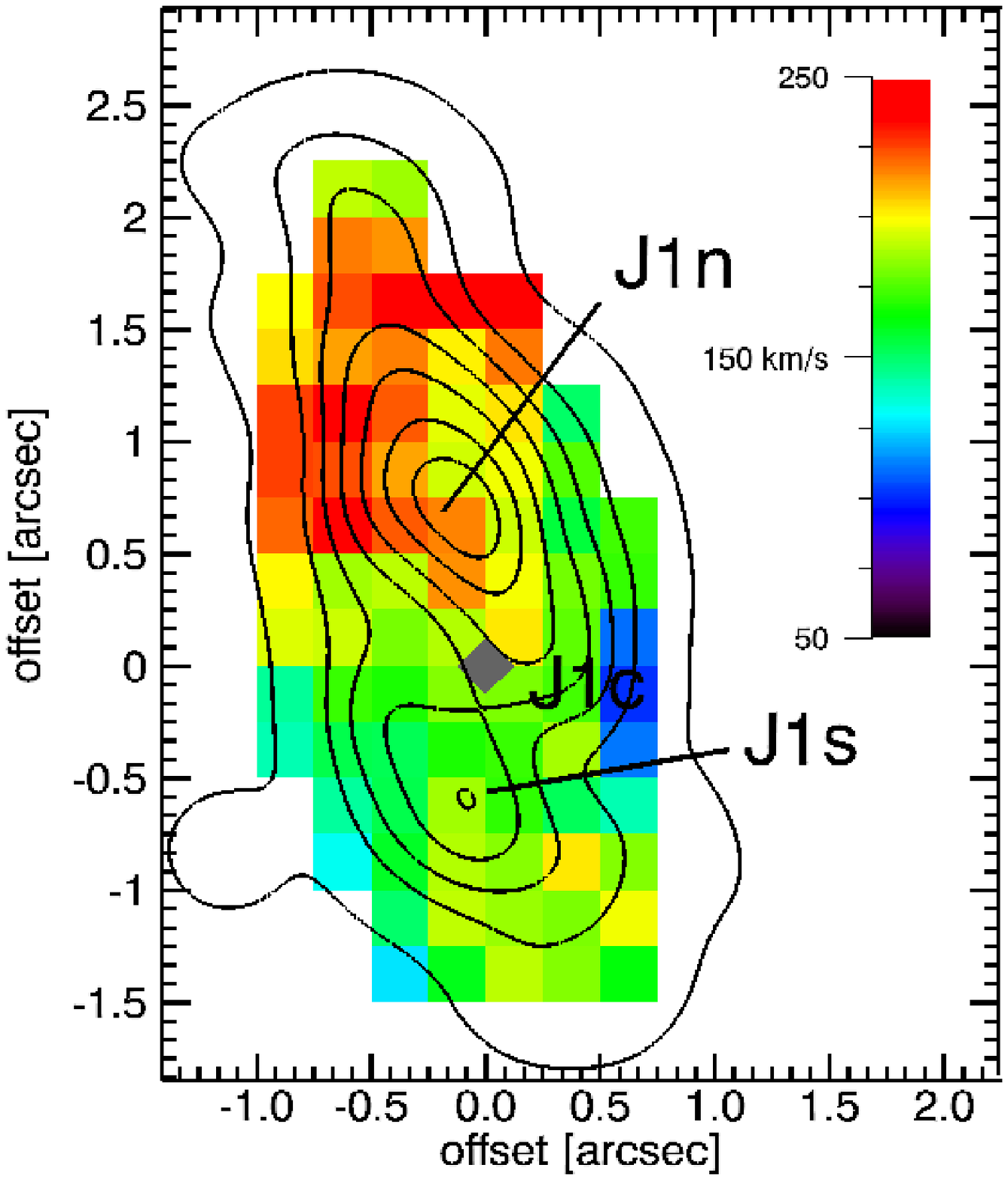}
\caption{Map of relative velocities {\it (top)} of \ha\ and full width
at half maximum of \ha\ {\it (bottom)}. In both plots, contours indicate
the K-band surface brightness with J1c removed, and the diamond
indicates the location of J1c. North is up, east to the left in each
panel.
\label{fig:velmap}}
\end{figure}

The SPIFFI K-band integral-field spectroscopy of J1 is of superb quality
and allows us to map the intrinsic kinematics within J1 using the spectral
positions of the \ha\ line cores (fitted with Gaussian line profiles)
across an area of $1.75\arcsec \times 3.25\arcsec$, with uncertainties of
$\lesssim 15$ \kms\ (Fig.~\ref{fig:velmap}). With a spatial resolution
of 0.4\arcsec$\times$0.6\arcsec\ in right ascension and declination,
respectively, corresponding to $\sim 4.4\times5.4$ seeing disks.  Hence,
the data set is well resolved spatially. Velocity dispersions of the
lines are $\sigma =$55$-$106 \kms\ (corrected for the instrumental
resolution).  Velocities vary by 190 \kms\ with the steepest overall
velocity gradient increasing from north-east to south-west. Although
this gradient is continuous, it is not strictly monotonic, and the
data suggest velocity shears along two axes, which coincide with the
regions of maximum K band surface brightness in the J1c removed data sets
(Fig.~\ref{fig:velmap}). The differences in the peak velocities in these
two dynamical regions are highly significant ($> 3 \sigma$) and are thus
not consistent with a single velocity gradient.

The close spatial alignment between star-formation (traced by the \ha\
surface brightness), gravitational potential (traced by the velocity map),
and stellar population (traced by the K band continuum and J$-$K color)
is intriguing (Fig.~\ref{fig:velmap}). This coincidence of the velocity
shear, star-formation, and K-band continuum can be naturally explained
if J1n and J1s are a close, likely merging, galaxy pair. Moreover, this
hypothesis is supported by our population synthesis fits discussed in
\S\ref{subsec:j1c}, which imply that J1n and J1s are dominated by stellar
populations of different ages and extinctions, and they have different
velocity dispersions (Fig.~\ref{fig:velmap}). At a projected distance of
$\sim 12 {\cal M}^{-1}$ kpc, the internal dynamics of J1n and J1s appear
to be driven by their own gravity.  The evidence for this is our finding
that both have velocity gradients which align well with their axes of
the most extended K-band continuum emission.  This indicates that the
central parts of either components are likely still dominated by their
individual potential wells.

\begin{figure}[hbt] 
\epsscale{0.7}
\plotone{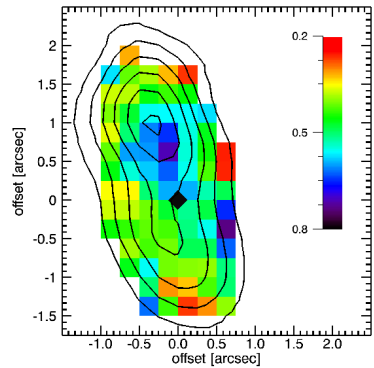}
\caption{Map of the [NII]\lam6583/\ha\ line ratios for the J1
complex. Contours indicate the distribution of \ha\ line surface
brightness. The grayscale bar on the right indicates the
values of the [NII]\lam6583/\ha\ ratios. North is up, east to the
left. The diamond marks the position of J1c.
\label{fig:n2map}}
\end{figure}

We have also mapped the [NII]/\ha\ line ratios (Fig.~\ref{fig:n2map})
over most of J1, where both lines are detected at $>5\sigma$
significance. Ratios are [NII]/\ha$\sim 0.3-0.4$ over most of J1s,
in the typical range of metal-rich H II regions and galaxy nuclei in
the nearby universe. [NII]/\ha\ ratios peak in J1n, about 0.8\arcsec\
to the north from J1c, and $\sim 0.6$\arcsec\ from the \ha\ peak. The
maximum ratio is [NII]/\ha$\sim 0.7$, which is significantly larger
than the ratios generally found in HII regions \citep[][and references
therein]{pettini04, osterbrock89}, and is more typical of starburst
nuclei in the local universe \citep[e.g.,][]{lehnert96a}.  By analogy,
this possibly indicates a strong shock component due to an outflow (see
\S\ref{sec:wind} for a detailed discussion). [OIII]\lam5007 and \hb\
are found (at lower signal-to-noise) near the [NII] surface brightness
peaks in J1n and J1s, consistent with this hypothesis.

\subsection{Relationship between Gas and Continuum Emission}
\label{subsec:gascont}

In both J1n and in J1s, the brightest \ha\ emitting regions are also
the regions with the reddest $J-K$ colors. Rest-frame equivalent
widths are $\sim 50-150$ \AA\ and are positively correlated with $J-K$
color in J1n and J1s, with significances better than ${\cal O}(0.01)$
in non-parametric tests (low values indicate a highly significant
correlation; Fig.~\ref{fig:jmink}). The positive correlation is a clear
indication that the reddening is related to star-formation in regions
with variable and high extinction rather than the age of the stellar
population. If the red colors indicated the age of the stellar population,
we would expect that the reddest regions of the galaxy in $J-K$ would
be the ones with the lowest \ha\ equivalent widths.

In the case of SMMJ14011+0252, however, the highest equivalent widths are
reached in the reddest areas, indicating that the color is indeed due to
variable and high extinction, rather than an age spread in the stellar
populations in both J1n and J1s. In addition, where the surface brightness
of line emission is intense enough to allow for a robust measure of
H$\beta$, the reddest regions also have high H$\alpha$/H$\beta$ ratios.
This supports the hypothesis that the color variation is more likely
due to reddening than to large differences in ages of the stellar
populations.  Our finding also implies, however, that significant
numbers of \ha\ photons escape, perhaps a sign for an irregular, clumpy
dust distribution. \cite{chapman04} reach a similar conclusion based on
the differences in structure in high resolution radio maps compared to
high resolution imaging in the rest-frame UV. It is also possible that
a favorably oriented starburst-driven wind (\S\ref{sec:wind}) is giving
a less obscured view to some of the starburst region.

\subsection{Population Synthesis Models and the
Role of J1c}\label{subsec:j1c}

As outlined in the Introduction, the nature of the optical and
near-infrared continuum peak J1c is difficult to constrain. It has a
nearly circular symmetry which led \citet{smail05} conclude that J1c
cannot be gravitationally lensed and is thus not at the redshift of
SMM14011$+$0252 \citep[and in agreement with][that some of absorption
lines in the Barger et al. spectrum are coincident with optical
absorption lines from a galaxy at z=0.25]{swinbank04}. Moreover, its
smooth, featureless light profile does not appear typical for a strongly
star-forming dusty galaxy at high redshift seen in the rest-frame UV. Both
arguments, although circumstantial, imply that J1c could be a foreground
object along the line of sight which is not physically related to the SMG.
Unfortunately, the observed wavelengths of some of the optical absorption
lines at the redshift of the foreground cluster A1835 are degenerate
with UV absorption lines at the redshift of the SMG \citep{swinbank04,
smail05}, so that the spectrum is not necessarily a unique constraint.

\begin{figure}[hbt] 
\centering \epsfig{figure=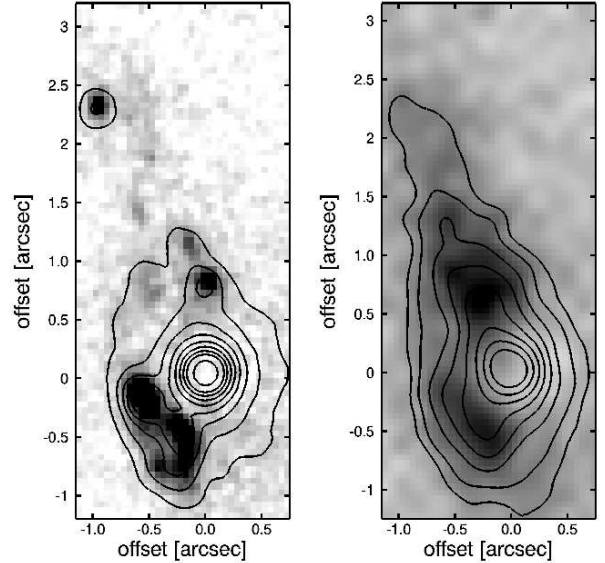,width=0.45\textwidth}
\caption{The light profile of J1c is very well fit with a Sersic
profile. The left panel shows the fit residuals in the F850 ACS image,
the right panel shows the residuals when subtracting the same, but
seeing-matched, model from the K band image of J1, adopting the same
pixel scale as in the ACS image. Contours indicate the original
surface brightness distribution of the F850LP and K band images,
respectively.  The morphology of J1n and J1s over the region of the
highest surface brightness region where J1c was located is an artifact
of scaling the model to the peak of the surface brightness of J1c in
each image.
\label{fig:j1crem}}
\end{figure}

We constucted an azimuthally averaged light profile of J1c from the F850LP
ACS image of SMMJ14011+0252.  To this one-dimensional profile, we fit a
Sersic profile convolved with the PSF (estimated using a nearby star in
the ACS image), obtaining a good fit with a Sersic model model of index
$n=1.25\pm 0.2$ and a half-light radius, $r_e=0.27\arcsec\pm 0.06$\arcsec.
The spatial resolution is $\sim 0.11$\arcsec.  We then subtract this
fit from the ACS image.  Fit residuals are $\lesssim 10$\%
in a 0.5\arcsec\ box aperture centered on J1c, and are likely due to
deviations in the core of the PSF from our simple Gaussian model and/or
extended emission from the sources J1s and J1n contaminating the fitted
light distribution of J1c. The left panel of Fig.~\ref{fig:j1crem}
shows the J1c removed F850LP ACS image.

We estimate the impact of J1c in each waveband by convolving our
best-fit Sersic model from the ACS image with the PSF appropriate for
each individual band and scaling to the measured peak brightness of
J1c.  This scaled, smoothed Sersic model is then subtracted from each
image centered on the position of J1c. Namely, we obtain magnitudes of
R$_{702} = 20.6$ mag, Z$_{850} = 20.5$ mag, J$ = 20.0$ mag, H$=19.0$
mag, and K$=18.2$ mag extracted from a 3\arcsec\ aperture centered on
J1c in the HST F702W WFPC2 and F850LP ACS images, and the ISAAC J, H,
and K band images, respectively. For the best-fit Sersic models to
J1c, integrated over 3\arcsec\ apertures, we find R$_{702} =21.5$ mag,
Z$_{850} = 21.3$ mag, J$=20.7$ mag, H$=19.9$ mag, and K$=19.2$ mag in
these bands. We use the fit residuals in each band to constrain
whether the light profile varies significantly with
wavelength. Overall fit residuals are $\le$15-20\% with random spatial
distribution, consistent with a light profile of J1c that is not a
strong function of wavelength (for illustration we use the K band data
in the right panel of Fig.~\ref{fig:j1crem}).  Our goal is to give an
upper limit on the overall impact of J1c on the submm source, which
might lead to an oversubtraction and is conservative.  However, doing
the subtraction this way will lead to a region centered on J1c that is
essentially at the background average in each image and appear as a
hole in the source (Fig.~\ref{fig:j1crem}).  Thus the detailed
morphology of J1n and J1s over the region of J1c should not be
overinterpreted (i.e., the hole in the light distribution is likely
not real and is not therefore a sign of strong shear for
example).  The S/N and rather coarse sampling in the SPIFFI data do
not allow a similar correction for the spectra (except in the case of
the H$\alpha$ equivalent widths that are enumerated in
Fig.~\ref{fig:eqwvsjmk}).

Deriving the magnitudes in the F702W, F850LP, J, H, and K bands based
on the scaled Sersic model, we fit the SED of J1c using the stellar
population synthesis package of \citet{bruzual03} for solar metallicity
and a Chabrier IMF. The SED is consistent with that of a z=0.25 galaxy
with an $\gtrsim$9 Gyr old stellar population of total mass $\lesssim
5\times 10^9$ \msun.  We do not find any suitable model when placing J1c
at $z\sim 2.6$, which supports the interloper hypothesis for J1c. For an
isothermal sphere and assuming that the projected half light radius is
the half mass radius of the galaxy, the stellar mass estimate corresponds
to a velocity dispersion which is consistent with $\sigma \la 55$ \kms\
adopted by \citet{smail05}.

From the data sets with J1c removed, we investigate the stellar
populations in J1s and J1n, assuming continuous star-formation histories,
solar metallicity, and a Chabrier IMF. Using only one star-formation
history and model is all that is justifiable given the considerable
uncertainties in removing J1c, and the overall faintness of J1n and
J1s. For J1s, our fits are consistent with a $\gtrsim 3\times 10^8$
Gyr old stellar population in J1s, $A_V\sim 0.8$ mag extinction, and a
mass of $\sim 8\times 10^9 {\cal M}^{-1}$ \msun\ (with the magnification
factor ${\cal M}$). For J1n, we find a $\lesssim 1\times 10^8$ yrs old
stellar population, A$_V\gtrsim 6$ mags, and a stellar mass of about
$3\times 10^{10} {\cal M}^{-1}$ \msun.  These age estimates are in
approximate agreement with what would be estimated from the values of
the rest-frame H$\alpha$ equivalent widths.  Fig.~\ref{fig:eqwvsjmk}
shows that the rest-frame equivalent widths of H$\alpha$ are typically
about 80 and 120 \AA\ for the most intense H$\alpha$ emitting regions
of J1s and J1n respectively.  Models of continuous star formation and
solar metallicity imply ages of a few 100 Myrs or less, with higher
equivalent widths corresponding to younger ages \citep{leitherer99}.
The H$\alpha$ to H$\beta$ line ratios imply significant extinction in
both J1n and J1s, as discusssed in \S\ref{subsec:bpt} and shown in
Table~\ref{tab:emissionlines}.

These mass estimates are likely only lower limits. To isolate the
individual components, we extract the photometry from square apertures
with 1.5\arcsec\ on each side (which is at least twice the seeing disk
in any of the images). From a sample of 20 SMGs and optically faint
radio galaxies with ACS imaging, \citet{smail04} find half-light
diameters D$_{1/2} \sim$ 1.5\arcsec- 3\arcsec. For ${\cal M}$=3-5 this
implies that we are underestimating the stellar mass by a factor $\sim
2-6$, depending on the exact morphology and light profile of J1n.
Moreover, our population synthesis fits might be biased by the strong
and clumpy extinction discussed in \S\ref{subsec:gas}.  Consequently,
we likely underestimate the overall age and extinction and hence the
stellar mass. This also raises worries that a significant fraction of
the total stellar population might be undetected due to strong
extinction, if the most strongly dust-enshrouded areas are optically
thick to the rest-frame R-band continuum (observed K band).  This
concern of course generically applies to photometric studies of dusty
high-redshift galaxies. Therefore analyzing spectra appears to be more
promising for estimating the physical properties of these dusty
systems.

We will in the following adopt $\ga 6\times 10^{10} {\cal M}^{-1}$
\msun\ as the stellar mass of J1n (correcting by a conservative factor
$\sim 2$ due to light not considered in the SED fit).  We warn that
given the relatively crude manner in which this estimate was made,
and the intrinsic difficulty in estimating the mass and star-formation
properties in heavily obscured galaxies from UV/blue optical SEDs,
our estimates are comparably uncertain.

\section{Characteristics of a High-redshift Starburst}\label{sec:maxburst}
\subsection{Star-formation Intensity}\label{subsec:sfintens}

Observations of low-redshift starburst galaxies suggest a fundamental
upper threshold for the star-formation rate in galaxies
\citep{lehnert96b, meurer97}. Such an upper limit may be indicative of
self-regulation processes limiting the gas collapse (and subsequent
star-formation) through the negative feedback of superwinds
\citep[e.g.,][see also \S\ref{sec:wind}]{lehnert96b} or perhaps
dynamical processes and disk instabilities \citep[e.g.,][]{meurer97}.
It is therefore interesting to investigate how SMMJ14011$+$0252 J1
relates to the ``maximal burst'' galaxies observed at low redshift.

For low-redshift star-forming galaxies, \citet{meurer97} propose
an upper bound of SFR$_{max} \sim 45$ \msun\ yr$^{-1}$ kpc$^{-2}$
within one half-light radius $r_{1/2}$. The exact value of the maximal
star-formation intensity depends on the choice of cosmology, the form
of the chosen initial mass function, the relative role of extinction,
the ability to measure half-light radii accurately at wavelengths where
a substantial fraction of the bolometric luminosity originates, etc., and
thus the numerical value of this threshold should not be interpreted too
strictly. However, it does provide a useful guide on when star-formation
becomes ``maximal'' -- whatever the cause \citep[see also][]{lehnert96a,
lehnert96b, tacconi06}. Do extreme stabursts in high-redshift galaxies
have similar intensities? We will in the following adrdress this question
empirically, comparing the star-formation intensity in SMMJ14011+0252 with
low redshift starbursts \citep{lehnert96b}, using the same techniques.
Such an estimate does not depend on the strength of the gravitational
lensing, because surface brightness is conserved and $\lesssim 10$\%
of the \ha\ emission is due to the wind component (\S\ref{sec:wind}),
which is negligible.

We estimate the half-light area in \ha, $A_{1/2,J1n}^{\ha}$, by
counting the pixels in J1n and J1s, respectively, that are above the
50\% percentile of the flux distribution \citep[following the method
of][]{erb03}. In J1n and J1s we find $A_{1/2}^{J1n}= 1.1 \square\arcsec$
(68 kpc$^2$), and $A_{1/2}^{J1s}=0.875\ \square\arcsec$ (56 kpc$^2$),
respectively.  Using the measured \ha/\hb\ decrements from these
regions to correct the \ha\ emission line fluxes for extinction,
we estimate intrinsic \ha\ fluxes, $F_{J1n}=1.1 \times\ 10^{-18}\
W\ m^{-2}$ in J1n and $F_{J1s}=0.37 \times\ 10^{-18}\ W\ m^{-2}$ in
J1s, respectively. We follow \citet{kennicutt98} with a 1-100 \msun\
Salpeter IMF to estimate star-formation rates, i.e. we adopt SFR = $7.69\
\times\ 10^{-35}\ {\cal L}_{H\alpha}\ M_{\odot}\ yr^{-1}$, where the \ha\
luminosity ${\cal L}_{H\alpha}$ is given in Watts.  Our adopted IMF gives
a total star-formation rate similar to that if we had adopted a more
appropriate IMF such as a Kroupa or Chabrier, namely 450 ${\cal M}^{-1}$
\msun\ yr$^{-1}$ in J1n and 150 ${\cal M}^{-1}$ \msun\ yr$^{-1}$ in J1s,
respectively.  We thus find star-formation rate densities of 7 \msun\
yr$^{-1}$ kpc$^{-2}$ in J1n and 3 \msun\ yr$^{-1}$ kpc$^{-2}$ in J1s,
respectively. In either case, the intensities are well below the limit
suggested by \citet{meurer97}.

Star-formation rates based on \ha\ luminosities tend to be lower by up
to about an order of magnitude compared to estimates based on infrared
luminosities.  This is because the most intense star-forming regions are
likely to be optically thick to dust at the wavelength of \ha, so that the
total star-formation rate will be under-estimated. Using the FIR-estimated
star-formation rate of Paper~I for a 1-100 M$_{\sun}$ Salpeter IMF, 1920
${\cal M}^{-1}$ M$_{\sun}$ yr$^{-1}$, and the measured half-light radius
of the \ha\ emission, we find $\Sigma_{FIR,H\alpha} = 28$ \msun\ yr$^{-1}$
kpc$^{-2}$. Using the FWHM of the CO emission line region to estimate
the diameter of the star-forming region \citep[1.6\arcsec$\times$0.5\arcsec
and 2.5\arcsec $\times$0.5\arcsec][]{downes06}, we find a similar
star-formation intensity of $\Sigma_{FIR,CO}= 48-30$ M$_{\sun}$ yr$^{-1}$
kpc$^{-2}$. Both estimates are within the range of the $80\pm 20$
M$_{\sun}$ yr$^{-1}$ kpc$^{-2}$ that \cite{tacconi06} found for a sample
of SMGs using high resolution CO observations. \citep[To mitigate against
the considerable uncertainties in calibrating star-formation intensities
from surface brightnesses, we note that the FIR luminosity of J1 from
Paper~I, $L_{FIR}=2.3 \times 10^{13} {\cal M}^{-1}$ L$_{\sun}$, and
the CO radius of \citet{downes06}, correspond to a surface brightness
of $S_{FIR} = 7.3-18\times 10^{10}$ L$_{\sun}$ in J1, compared to a 90\%
percentile of the effective surface brightness, $S_{e,90}=2.\times10^{11}$
L$_{\sun}$ kpc$^{-2}$, given by][]{meurer97}.

In spite of the large uncertainties of each method, none of the
estimates for J1 exceeds the star-formation intensities in low
redshift ``maximal burst'' galaxies.  

The most fundamental physical motivation to posit a stringent upper
bound to the  star-formation rate in a self-gravitating system comes
from causality. This limit is given by consuming all of the gas (modulo
a star-formation efficiency per unit gas mass) within one dynamical time,
either a crossing or free fall time scale \citep[see,
 e.g.,][]{lehnert96b}. \cite{tacconi06} find that none
of the SMGs, including SMMJ14011$+$0252, violate causality
arguments.  Specifically, within the context of such a hypothesis,
\citeauthor{tacconi06} find that SMGs form stars with efficiencies similar
to those seen in local galaxies and star-forming regions, 0.1-0.3,
and within a few dynamical timescales.  Our results here suggest that
SMMJ14011$+$0252 does not violate the stringent causality limits either.

\subsection{No ``new mode'' of star-formation}
\label{subsec:newmode} 

At low redshift, only ULIRGs reach similarly large infrared
luminosities as SMGs, $L\gtrsim 10^{12}$ L$_{\odot}$, whose properties
are typically attributed to the effects of massive gas collapse during
major mergers \citep{sanders96}. Motivated by the extended nature of
star formation in SMMJ14011$+$0252, \citet{goldader02} hypothesize
that there might be an intrinsically more efficient mechanism of star
formation at high redshift, with no equivalent in the local universe,
which does not require a major merger to trigger extreme star
formation. 

Our data do not confirm this hypothesis, in fact, they rather point
towards SMMJ14011+0252 being a merger (\S\ref{subsec:gas}). As discussed
previously, our data do not indicate a higher intensity of star
formation than at low redshift, and the half-light radii of \ha\
emission are similar to those of low-redshift ULIRGs, correcting for
gravitational magnification of SMMJ14011$+$0252. For example, the
largest \ha\ half-light radius we find in the data set of
\citet{colina05}, in the same way as for SMMJ14011$+$0252 (see above),
is 2.1 kpc, compared to $\sim 1.7$ kpc for J1n (for a magnification
${\cal M}= 5$). We conclude that we do not find evidence for a
peculiar, more efficient ``high-redshift'' star-formation mode in
SMMJ14011$+$0252.

\section{A Starburst-driven Superwind at $z=2.6$}
\label{sec:wind}

\subsection{Properties of the Superwind}

Actively star-forming galaxies with intensities exceeding 0.1 \msun\
yr$^{-1}$ kpc$^{-2}$ are known to drive ``superwinds'', irrespective
of redshift \citep{heckman03}. SMMJ14011$+$0252 J1 easily surpasses
this limit (\S\ref{sec:maxburst}) and can therefore be expected to
drive such a wind. Blue asymmetries in emission line profiles,
e.g. \ha\ and [NII]\lam6583 line profiles are common in starburst
galaxies in the local Universe with substantial evidence for driving
vigorous outflows \citep{lehnert96a}, and are also the most common
evidence for winds at high redshift. Velocity offsets of a few 100
\kms\ in UV absorption lines tracing the ISM relative to the systemic
redshift are also frequently used as evidence for winds
\citep{shapley03,swinbank05}. In J1, rest-frame UV absorption lines
are blueshifted relative to \ha\ by $\sim 500$ \kms\
\citep{ivison00,tecza04}, and we also observe blue wings of \ha\ and
[NII]\lam6583 emission lines in J1n (see the inset of
Fig.~\ref{fig:kspec}). Gaussian fits to the residuals have relative
offsets of $\sim -330$ \kms\ between line wing and core for the \ha\
lines and $\sim -350$ \kms\ for [NII] in J1n. About $10$\% of the \ha\
flux is in the wing, and $\sim 25$\% of [NII]\lam6583. J1s does not
have a pronounced blue wing.

The wavelength-coverage and quality of our J1 data sets are
outstanding for a strongly starforming galaxy at $z\sim 2.6$, 
allowing for a more detailed analysis to firmly establish that
starbursts and outflows in high redshift galaxies are causally linked
by the same basic mechanism as at low redshift. Starburst-driven winds
at low redshift are caused by thermalized ejecta of supernovae and
stellar winds producing an over-pressurized, expanding bubble of hot,
X-ray emitting gas, which sweeps up, entrains, accelerates, and
ionizes the ambient interstellar medium \citep{heckman90}. This
results in high gas pressures and shock-like optical emission line
ratios.  Positive correlations between ionization state and line
widths have been observed at low-redshift
\citep[e.g.][]{lehnert96a,rupke05}, implying that the relative
importance of shock heating increases as the gas is accelerated to
outflow velocities of a few hundred \kms, where the emission line
luminosity rapidly increases with increasing shock speed
\citep{dopita95}.

\begin{figure}[hbt]
\centering
\epsfig{figure=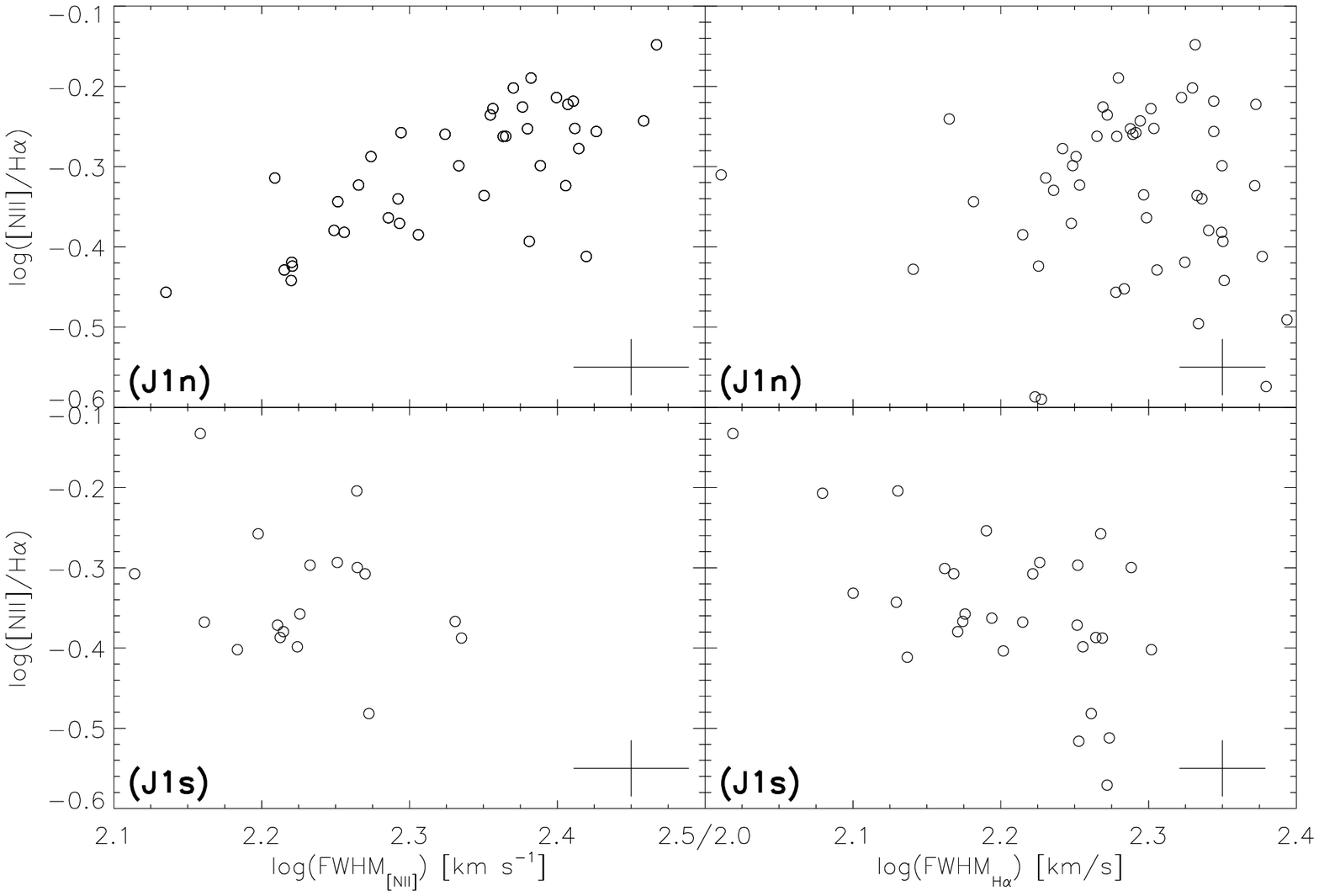,width=0.45\textwidth}
\caption[Line ratio / FWHM relationships in J1n and J1s.]
{[NII]\lam6583/\ha\ line ratio as a function of [NII] and \ha\ FWHM
for J1n {\it (top, right and left panels)} and J1s {\it (bottom, right
and left panels)}. [NII] FWHM and line ratio are correlated in J1n, but
not in J1s. For the \ha\ FWHM, no similar correlation is found in J1n
or J1s. Interestingly, the correlation seen in the [NII]\lam6583/\ha\
line ratio and [NII] FWHM for J1n is virtually identical in the
range of values and the correlation to that observed in a sample
of local starburst galaxies exhibiting superwinds \citep[][their
Fig.~17]{lehnert96a}. Crosses indicate typical uncertainties.}
\label{fig:widthcorrelj1n}
\end{figure}

For J1 we can investigate each of these properties explicitly.
[NII]\lam6583/\ha\ line ratios are $\sim 0.7$ in the integrated
emission of J1n, which is generally higher than observed in HII regions
(photo-ionization by massive stars), and can be easily caused by
shocks. The [NII]/\ha\ line ratio in the wings of J1n is 0.83 indicating
a higher shock contribution in the outflowing gas.  We also find a
good correlation between [NII]\lam6583/\ha\ ratio and [NII]\lam6583
line width in J1n (left panel of Fig.~\ref{fig:widthcorrelj1n}), but
not in J1s. We also find no correlation with \ha\ line widths (right
panel of Fig.~\ref{fig:widthcorrelj1n}), similar to nearby ``superwind''
galaxies \citep{lehnert96a}.

With the electron densities derived in \S\ref{subsec:bpt} we estimate
the pressure of the electron gas using the conversion given in
\citet{lehnert96a} and considering that much of the [SII] and [OII]
lines fluxes are produced in partially ionized zones of the nebulae
\citep{sm79}: $P \approx 4 \times 10^{-12}\ n_e\ {\rm dyn}\ {\rm
cm}^{-2}$, or P$\approx 2\times10^{-9}$ dyn cm$^{-2}$ for the electron
densities $\sim 400$ cm$^{-3}$ derived in \ref{subsec:bpt}.  This is in
excellent agreement with the pressures in the nuclei of nearby starburst
galaxies with evidence for driving superwinds \citep{lehnert96a}, and
is factors $\sim 10^{3-4}$ higher that the pressure of the local ISM in
the Milky Way, providing the necessary prerequisite for an expanding or
outflowing gas bubble.

However, compared to low-redshift starburst galaxies, the wind in
J1 seems to have a relatively high surface brightness. Maximum \ha\
surface brightnesses in the low redshift sample of \citet{lehnert96a} are
$\lesssim 800\ L_{\odot}$ pc$^{-2}$ and for our cosmology would correspond
to an observed flux $f_{obs} = 2.3 \times 10^{-21} h_{70}^{-2}$ W m$^{-2}$
pix$^{-1}$ at the redshift of J1, about an order of magnitude lower than
the total \ha\ flux in the wing of J1 ($\sim 10$\% of the total \ha\
flux, or $\sim 1.1 \times 10^{-19}$ W m$^{-2}$ pix$^{-1}$). This might be
due to a higher gas fraction in J1, leading to a larger covering factor
of gas clouds that are being shocked. \citet{tacconi06} estimate a gas
fraction of $f_{gas}\sim 0.4$ in SMGs supporting such speculation.

\begin{figure}[htb]
\centering
\epsfig{figure=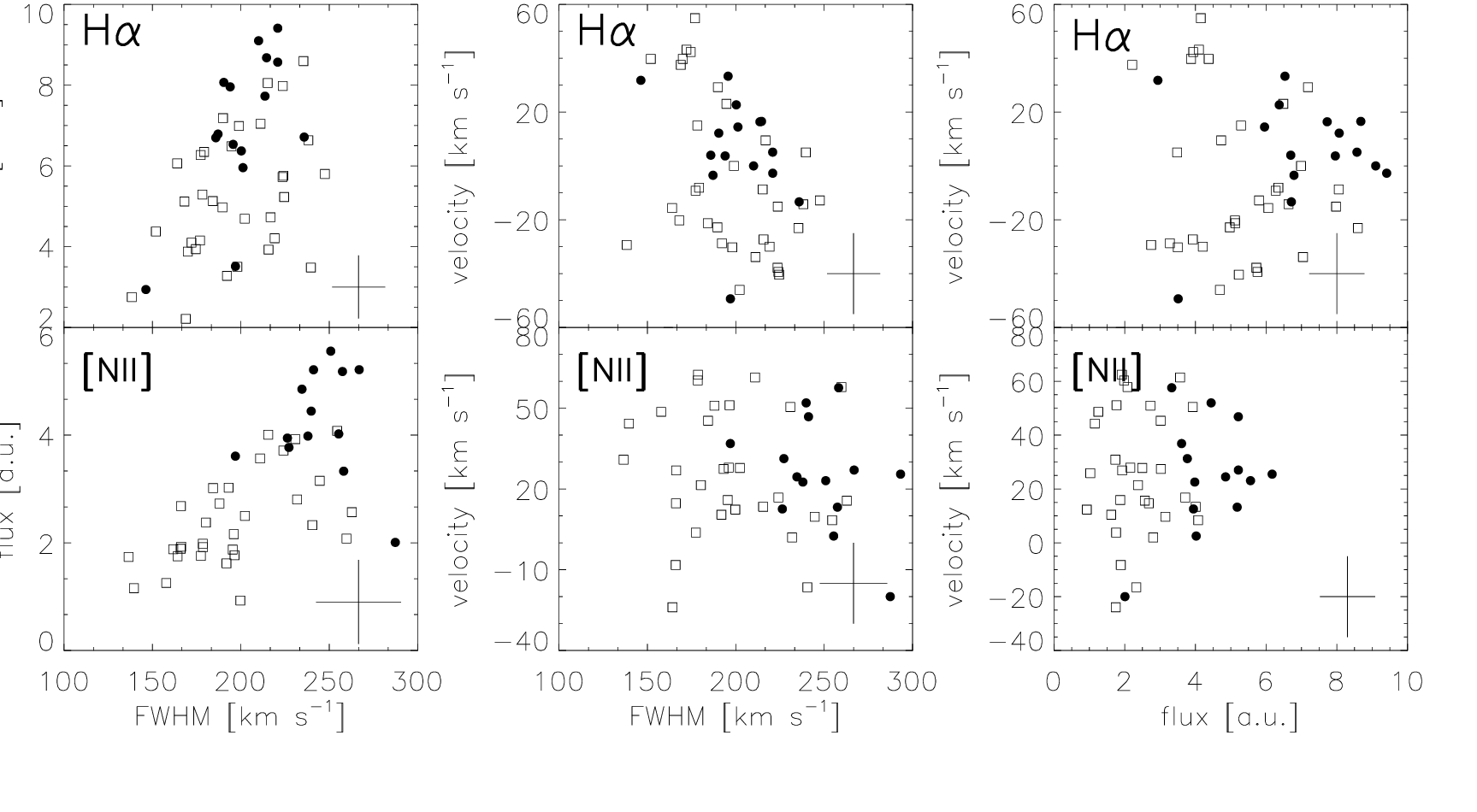,width=0.28\textwidth,angle=90}
\caption[Emission line correlations for NII]
{Emission line correlations for \ha\ and [NII]\lam6563 in J1n. In the
upper and lower panel of each column, we show the correlations for \ha\ and
[NII]\lam6583 respectively.  Open squares in each diagram represent
pixels with low ionization, [NII]/\ha$< 0.55$, whereas filled circles
indicate pixels with [NII]/\ha$> 0.55$. Crosses indicate typical
uncertainties.
\label{fig:smm14.correls}}
\end{figure}

Since the wind causes a correlation between [NII]\lam6583/\ha\ ratio
and the [NII] line width, we also investigate its impact on the overall
velocity field (Fig.~\ref{fig:velmap}), again using correlations between
the spatially resolved emission line properties to localize and compare
the signatures of mechanical heating and kinematic parameters.  The upper
and lower panels of Fig.~\ref{fig:smm14.correls} show the correlations
between line properties for the H$\alpha$ and [NII]$\lambda$6583 lines
in J1n, respectively. The data are given separately for spatial pixels
with high and low [NII]/\ha\ ratios. Empty squares indicate spatial
pixels where [NII]/\ha$<0.55$, filled dots show spatial pixels where
[NII]/\ha$>0.55$. Line widths and fluxes are correlated for [NII], tracing
the stronger contribution of mechanical heating in the blueshifted [NII]
gas, but none of the other line properties appear to be correlated.
In particular, the position of the line centroids are not correlated
with flux or line width (middle and right panels), suggesting they
are dominated by large-scale gravitationally-driven motion similar to
low-redshift starburst galaxies \citep[e.g.,][]{lehnert96a}. This is the
case for pixels with high and low [NII]/\ha\ ratios.  Fully consistent
with the integrated spectrum (Paper~I), this shows that gas ionization
is dominated by the ionizing photons from the intense star-formation,
and that the velocities traced by the \ha\ line centroids are not overall
significantly altered by the wind.

\subsection{Self Regulated Star-Formation?}
\label{subsec:selfreg}

As discussed previously, there are several arguments about what might
be regulating star formation within galaxies \citep[see also, e.g.,
][]{tacconi06,lehnert96b}, including negative feedback from the superwind,
if the overpressureized bubble can plausibly provide pressure support. Can
this hypothesis explain the variation of velocity dispersions in J1,
and limit the rate of star-formation?  Adopting a disk geometry, this
scenario implies that gas collapse can be balanced if the momentum flux
injection, ${\cal P}_{wind}$, is comparable to the midplane pressure,
${\cal P}_{mid}$. In hydrostatic equilibrium, ${\cal P}_{mid}$
is,

\begin{eqnarray}
\nonumber
{\cal P}_{mid}= \frac{1}{2}\pi G \Sigma_{tot} \Sigma_{gas} = 4.6
\times 10^{-10}\ \Sigma_{tot,9}\ \Sigma_{gas,8}\ {\rm dyn}\ {\rm 
cm}^{-2} 
\end{eqnarray}

where $\Sigma_{tot,9}$ and $\Sigma_{gas,8}$ are the surface mass
densities of all of the matter and just the gas in units of $10^9$
M$_{\odot}$ kpc$^{-2}$ and $10^8$ M$_{\odot}$ kpc$^{-2}$,
respectively. We use the size of molecular emission
\citep[FWHM$=$2.2\arcsec$\times$0.5\arcsec][]{downes03}, the molecular
gas mass estimated in Paper I (and references therein), and our
dynamical mass estimate, M$_{dyn,J1}\sim 10^{11}$ \msun (see
\S\ref{sec:dynmass}), to estimate a total midplane pressure, ${\cal
P}_{mid}\approx10 \times 10^{-9} {\cal M} $ dynes cm$^{-2}$. The
pressure provided by the wind can be derived from the total momentum
flux of the outflow. Following \cite{heckman90} and \cite{lehnert96a}
we parametrize the momentum injection by the wind as
$\dot{p}_{winds}\approx3\times10^{34}{\cal L}_{\rm IR,11}$ dynes, where
${\cal L}_{\rm IR,11}$ is the infrared luminosity in units of 10$^{11}$ \Lsun.
With the infrared luminosity estimated in Paper~I, 2.6$\times$10$^{13}
{\cal M}^{-1}{\cal L}_{\sun}$, this implies a momentum flux injection
rate of 8$\times$10$^{36} {\cal M}^{-1}$ dynes, similar to the most
powerful local starbursts \citep{heckman90}. With the above size
estimate, we find ${\cal P}_{wind}\approx$15$\times$10$^{-9}$ dynes
cm$^{-2}$. This implies ${\cal P}_{wind} \approx 0.3 {\cal P}_{mid}$
and ${\cal P}_{wind} \approx 0.5 {\cal P}_{mid}$ for ${\cal M} =5$ and
3, respectively. Although this is somewhat smaller than unity if taken
at face value, given the large uncertainties this is clearly
consistent with the notion that star formation may be self-regulated
through its own feedback in the most powerful starbursts, in
particular at high redshift.

\section{The Blue Component J2 -- Intrinsic Properties and a Probe for
the Dynamical Mass of J1}
\label{sec:j2}

Our data sets also include component J2 of SMMJ14011$+$0252, which is
about 1.3\arcsec\ to the north-west from J1c. J2 is considerably bluer
than J1 at observed optical wavelengths \citep[e.g.,][]{ivison00}.
\citet{smail05} pointed out that its rest-frame UV colors are similar
to those of $z\sim3$ Lyman-break galaxies, although its redshift is
relatively low for such a comparison. Our population synthesis fits
(see \S\ref{subsec:j1c}) indicate a $\gtrsim 3\times 10^8$ yr old
stellar population with $ 4.5\times 10^9 {\cal M}^{-1}$ \msun\ stellar
mass and $A_V = 0.5$ mag. This implies an older and more massive stellar
population than favored by \citet{motohara05}, although it is nonetheless
a consistent result, because \citeauthor{motohara05} only compared their
photometry with an instantaneous burst and a $10^9$ yrs old continuous
star-formation episode.

As noted by previous authors \citep[e.g.,][]{ivison01, tecza04}, the
spectrum of J2 is blueshifted by $-160 \pm 18$ \kms\ with respect to
J1. In addition to detecting \ha, we also identify [NII]\lam6583 line
emission in the K band, and the [OIII]\lam\lam4959,5007 doublet in the H
band. \hb\ was not detected in J2, probably because of a nearby strong
night sky line (see Table~\ref{tab:emissionlines} for the emission
line properties).

These results differ from those of \citet{motohara05}, who identified
\hb, but not [OIII]. The discrepancy is easily explained by the lower
resolution of their data ($R=210$) and unfortunate spectral positions of
the emission lines (\hb\ is near a strong night-sky line, [OIII]\lam5007
lying within a set of a strong telluric absorption features).  We measure
a $3\sigma$ limit on the \hb\ flux, $F_{H\beta} \leq 9\times 10^{-21}$
\intfluxmks, compatible with relatively low extinction as indicated by the
blue, ``LBG-like'' colors of J2 \citep[see e.g.][]{smail05}, $E(B-V)\sim
0.23$, and an extinction corrected (lensed) \ha\ luminosity $L_{H\alpha}
= 2.3\times 10^{35}$ W, which corresponds to a star-formation rate of
$\sim 18 {\cal M}^{-1}$ \msun\ yr$^{-1}$. For a magnification ${\cal M}
\gtrsim 3$, the star-formation rate is $\lesssim 6$ \msun\ yr$^{-1}$,
lower than the typical rates in UV-selected, star-forming BX galaxies
at similar redshift \cite[e.g.][]{erb03}, and two orders of magnitude
less than the highest estimates of the star-formation rates in J1.

\ha\ emission in J2 extends over $\sim1.25\arcsec \times 0.75\arcsec$
and is offset from the continuum peak by $\sim 0.4$\arcsec\ to the
north-east. Extracting both line and continuum information from the
same data cube and the reasonable signal-to-noise of both the line and
continuum emission imply that this offset is significant. We do not
measure a significant velocity gradient in the H$\alpha$ line emitting
gas, but measure an intrinsic FWHM$=66\pm 8$ \kms, for a reasonable set
of assumptions about the mass distribution, corresponds to a dynamical
mass of M$_{dyn,J2} \sim 10^9$ \msun.

\begin{figure}[htb]
\epsscale{1.0}
\plotone{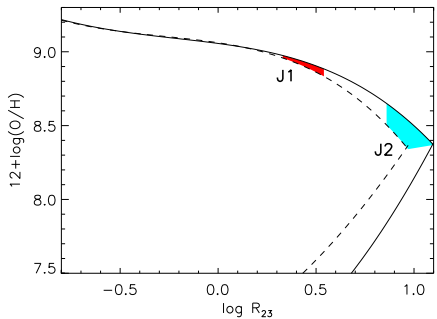}
\caption[\ha+NI spectra in J1.]
{An estimate of the Oxygen abundances using the R$_{23}$ method for J1 and
J2. Solid and dashed lines indicate the calibration for high and low
ionization, respectively.  The red and aqua colored regions show the
best estimate of 12+log(O/H) for J1 and J2 respectively and the ranges
shown indicate the 1$\sigma$ uncertainties in R$_{23}$.  The values of
12+log(O/H) are double valued, but the relatively high [NII]\lam6583/\ha\
ratio of both J1 and J2 suggest they lie on the upper branch.  For J2,
\hb\ was not detected, and we used the 3$\sigma$ upper limit on \hb\
to determine the abundance in J2.  Weaker \hb\ moves R$_{23}$ to even
higher values and only marginally lower 12+log(O/H) due to J2 already
lying near the upper envelope of R$_{23}$ values.}
\label{fig:metallicities}
\end{figure}

We calculate the [O/H] abundance of J2 from R$_{23}$ to compare with
the J1 abundance discussed in Paper~I and to investigate whether the
two components have similar or distinct evolutionary histories. As seen
in Fig.~\ref{fig:metallicities}, oxygen abundances are significantly
different in J1 and J2, using the 3$\sigma$ upper limit on \hb\ to
determine the abundance in J2. The [NII]/\ha\ ratio of $\sim 0.23\pm0.07$
in J2 indicates that the upper branch is appropriate for estimating
the metallicities.

The position of J2 in the emission line diagnostic diagrams of
\citet{osterbrock89} is also interesting.  The [SII] line doublet is not
detected in J2. The upper limit on the \hb\ flux only provides a lower
limit for the [OIII]/\hb\ ratio. The emission line diagnostic ratios
and excitation diagrams indicate comparably high ionization for J2,
placing it near the limit of the AGN portion of the diagram. The distinct
position of J2 indicates higher ionization or temperature compared to
J1, and is likely a direct consequence of the lower metallicity of J2
(and perhaps a concomitant lower dust content).

The orbit of J2 can also be used to estimate an approximate dynamical mass
of J1, which is particuarly valuable, as the complex velocity field in J1
makes it difficult to robustly estimate a dynamical mass estimate from
the intrinsic kinematics. Following Paper~I we add $6.5 \times 10^{10}\
{\cal M}^{-1}$ \msun\ in gas to the $\gtrsim 6 \times 10^{10} {\cal
M}^{-1}$ \msun\ of stellar mass we found in \S\ref{subsec:j1c}. J1 appears
to dominate the overall baryon budget by about an order of magnitude,
which is consistent with the narrow line widths in J2 compared to J1n,
and is certainly suggested by our astrometry which places the peak of the
CO emission within the emission of J1n.  Placing the barycenter on J1n, we
use a simple virial estimate for the mass, $M_{\rm J1}^{\rm dyn} = \Delta
V^2\ R\ G^{-1} = 2.\times 10^{10}\ \Delta V^{\rm J1J2}_{\rm 100}\ R^{\rm
J1J2}_{\rm kpc}\ \msun \sim 1.0\times 10^{11}$ \msun, where G is the
gravitational constant, $\Delta V^{\rm J1,J2}_{\rm 100}$ is the relative
velocity between J1 and J2 in units of 100 \kms, and $R^{\rm J1J2}_{\rm
kpc}$ is the distance between J1 and J2 in kpc. Since true inclination,
eccentricity, form of the potential, etc., are all unconstrained and
the physical separation is unknown, the true mass is likely factors of a
few larger (both radius and velocity for example are seen in projection).

\section{Surface Mass Densities and the Future Evolution of SMMJ14011+0252
and other SMGs} \label{sec:dynmass}

In addition to the high dynamical masses of SMGs \citep{genzel03,
greve05}, mass densities \citep{tacconi06}, their luminosity-weighted
ages, high star-formation rates, and possibly strong clustering indicate
that they will likely evolve into massive early-type galaxies in cluster
environments at low redshift \citep[][and references therein]{smail04}.
Interestingly, the characteristics of local galaxies also seem
highly dependent on their mass surface densities (and perhaps less
so on their overall mass), as suggested by a recent analysis of
\citet{kauffmann06}. They studied nearly 400,000 low-redshift galaxies
drawn from the Sloan Digital Sky Survey (SDSS), and find significant
differences between the structural parameters of early and late type
galaxies which does not strongly depend on the total mass of the
galaxy. Namely, they find that concentration parameters $C>2.5$ and
stellar mass surface densities $\mu_{\star}>3\times 10^{8}$ M$_{\sun}$
kpc$^{-2}$ correspond to the regime of galaxy spheroids and bulges,
independent of the total stellar mass of these systems.  Above this
threshold of mass surface density, they find that star formation is
increasingly suppressed (the specific star formation rate is low) and
must have ceased many Gyrs ago. As a consequence, \cite{kauffmann06}
hypothesize that with increasing compactness and surface density of
the galaxy, stars were formed in short, vigorous episodes at high
redshift, with extended periods of inactivity \citep[][parameterize
this in terms of a consumption time scale, t$_{cons}\propto
\mu_{\star}^{-1}$]{kauffmann06}. Substantial growth at later epochs was
then only possible through mergers of galaxies with low gas fractions. To
identify SMM14011$+$0252 and other SMGs as ``spheroids in formation'',
it is therefore not sufficient to address their large masses or even their
dynamical mass densities given their high gas fractions and unknown dark
matter distributions. They must also have high stellar mass densities,
short gas consumption time scales, and strong feedback suppressing
further star-formation. Do SMGs have all these properties?

Our analysis of the stellar mass surface density in SMMJ14011+0252 J1
is that it is at the transition value of $3\times 10^8$ \msun kpc$^{-2}$
and the dynamical estimate of the mass surface density in SMMJ14011+0252
J1 is $3\times 10^9$ M$_{\sun}$ kpc$^{-2}$, well above the transistion.
The stellar masses ($\langle M_{stellar}\rangle =3\times 10^{10}
M_{\sun}$) and average half-light radius ($\sim 5$ kpc) of a large sample
of SMGs \citep[e.g.,][]{smail04}, suggest average stellar mass surface
density of $\mu_{\star}^{SMG}= 4\times 10^8$ M$_{\sun}$ kpc$^{-2}$.
Similar estimates can be made from dynamical mass estimates and sizes
from CO interferometric observations \citep[$\sim$10$^9$ M$_{\sun}$
kpc$^{-2}$ for stars, gas, and possible contributions from dark matter;
][]{tacconi06, greve05}.  It appears that SMMJ14011+0252 and SMGs in
general have sufficiently high mass surface densities to be above the
critical point in the study of \cite{kauffmann06}.

Population synthesis fits of the stellar population of SMM14011$+$0252
yield ages of up to a few 100 Myrs, similar to the ages of typical
radio detected SMGs with spectroscopic redshifts \citep[$\sim$ few
100 Myrs;][]{smail04}. Based on the infrared luminosity and the gas
mass, we estimate a gas consumption time scale of about 30-40 Myrs
for SMM14011$+$0252. From a larger sample of SMGs with CO detections,
\cite{greve05} set a limit on the gas consumption time scale of $\ga$40
Myrs, consistent with that for SMM14011$+$0252.  Thus individual SMGs
appear to be forming most of their stars in intense bursts of moderate
duration (a few 100 Myr).  In addition, their star-formation appears to
be highly ``bursty'' independent of the duration of the star-formation.
The duty cycles of SMGs have been estimated through a variety of
methods to about $\sim 0.1$ \citep{chapman05, bouche05, genzel03,
tecza04, blain04}.   Again, SMGs seem to form their stars in intense
bursts of modest durations with long periods of relative inactivity.
To explain the fractions of local galaxies with strong 4000\AA\ breaks,
\citet{kauffmann06} infer gas consumption time scales of $\approx$100
Myrs, similar to estimates of SMGs.

As discussed in \S\ref{sec:wind}, SMM14011$+$0252 is driving a vigorous
superwind.  While the importance of starburst driven winds to galaxy
evolution is generally agreed upon, they are not thought to be powerful
enough or to accelerate material to the escape velocities of the most
massive galaxies \citep{heckman00}.  However, SMM14011$+$0252 is driving
a wind, which at the very least will lower the overall star-formation
efficiency of SMM14011$+$0252 (see discussion in \S\ref{subsec:sfintens}
and \S\ref{subsec:selfreg}).  If this is indeed a general characteristic
of SMGs, then the role of winds in their ultimate evolution could be
substantial.

Hence, the ``submillimeter bright phase'' of galaxies is characterized by
high surface mass densities, above the interesting dividing point of the
characteristics of local galaxies as found by \citet{kauffmann06}. Their
star formation appears to be episodic (duty cycle of $\approx$0.1) and
of relatively short duration \citep[few 100 Myrs, e.g.,][]{smail04},
and they have evidence for feedback which we have suggested may regulate
the intensity of their star-formation.  This adds further evidence that
SMGs do indeed have all the characteristics necessary to become local
massive early type galaxies.

\section{Summary}\label{sec:conclusions}

We presented an integral-field study of the rest-frame optical emission
line gas in the $z\sim 2.6$ SMMJ14011$+$0252 J1/J2 complex, allowing
unprecedented insight into the nature of a high-redshift starburst and
its outflow. Identifying J1c as a z$\sim$0.25 interloper through Sersic
profile and SED fitting, we removed the seeing-matched J1c contamination
from the images, and find that J1n and J1s appear as individual components
in all wavebands, at a projected distance of a few kpc. The positions
of these two components are in excellent agreement with the distribution
of \ha\ emission in our SPIFFI \ha\ map and also with the \ha\ velocity
field, which is reminiscent of two nearby co-rotating disks which are
marginally resolved spatially. Including J2, SMMJ14011+0252 thus appears
as a triple system. From the J2 orbital motion, we estimate a dynamical
mass of $M_{dyn,J1}\sim 1.0\times 10^{11}$ \msun\ for the J1n/J1s complex.

The dust-enshrouded J1n is the most massive component, as indicated by
its stellar mass, $\sim$few $\times 10^{10}$ \msun\ (compared to $\sim
1-2\times 10^{9}$ \msun\ for J1s and J2), and the bright CO line
emission, which coincides with J1n within the astrometric
uncertainties. The starburst in J1n is ``maximal'' with similar
intensity ($\lesssim 50$ \msun\ yr$^{-1}$ kpc$^{-2}$) comparable to
the apparent limit at low redshift. The \ha\ half-light radius is
similar to the \ha\ half-light radii in low-redshift ULIRGs. Overall,
star formation in J1n is intense, but its properties do not greatly
differ from low-redshift ULIRGs, highlighting the similarity between
SMGs and ULIRGs at optical wavelengths, i.e., in the extended
gas. Given these similarities and the complex large-scale kinematics,
J1 does not appear to be a good candidate for an alternative, highly
efficient ``high-redshift mode'' of star-formation, but appears
governed by similar rules as low-redshift galaxies with intense
star-formation, although it may be ``scaled up'' \citep{tacconi06}.

The intense starburst in J1 drives a superwind, as evident from blue
emission line asymmetries, offsets between rest-frame UV interstellar
absorption lines relative to \ha, and enhanced [NII]\lam6583/\ha\ line
ratios in J1n. The [NII]/\ha\ ratios correlate with [NII] line width,
indicating an increasing contribution of mechanical heating as the gas
is accelerated in the wind, similar to low redshift starburst-driven
winds. Measured densities (from the [OII]\lam\lam3726,3729,
[SII]\lam\lam6717,6731 doublets) indicate pressures of $\sim 2\times
10^{-9}$ dynes cm$^{-2}$, similar to pressures estimated in the expanding
bubbles of over-pressurized hot gas in local starburst galaxies. These
results are a direct indication that the basic physics of starburst-driven
winds are rather similar at low and at high redshift, and supports the
approach of studying local starburst galaxies to better understand the
basic mechanisms of high-redshift galaxy formation.  The strong wind
may explain why \ha\ equivalent width and reddening in SMMJ14011+0252
are positively correlated, which is likely a sign of patchy extinction.

The bluer component of the SMMJ14011$+$0252 system, J2, is very different
from J1, with mass $\sim 10^9$ \msun\ estimated from the narrow emission
lines and SED fitting. Its gas-phase oxygen abundance, measured from
R$_{23}$, $12+[O/H]= 8.5$, about 0.4 dex lower than in J1 (Paper~I).
This signals that the two galaxies have had independent evolutionary
histories, and that J2 is likely going to be accreted by the more
massive J1n.

The ``submillimeter bright phase'' of galaxies, which SMM14011$+$0252
is now in, is characterized by high surface mass densities, their star
formation appears to be episodic and of relatively short durations, and
they have evidence for feedback.  These are just the characteristics
needed to form early type spheroid dominated galaxies in the local
Universe \citep{kauffmann06}. We find that only J1n has the necessary
stellar mass surface density $\mu_{\star} > 3\times 10^8$ M$_{\sun}$
kpc$^{-2}$, while the blue components appear less concentrated. With
their current star-formation time scales, however, it appears unlikely
that they are likely to substantially change the final characteristics
of the most massive component J1n.

\acknowledgements
We would like to thank I. Smail and D. Downes for interesting discussions
on the properties of submm galaxies and SMMJ14011$+$0252 in particular and
D. Downes for providing mm maps of SMMJ14011$+$252 from his previous work
and some in advance of their publication.  We also thank R. Ivison for
providing the radio map of SMMJ14011$+$0252 and its surroundings which
proved crucial in obtaining the accurate astrometry presented here.
We further thank him, Ian Smail, and Scott Chapman for enlightening
discussions about the properties and nature of submm-selected galaxies.
AJB acknowledges support as a Jansky Fellow from the National Radio
Astronomy Observatory, which is operated by Associated Universities,
Inc. under cooperative agreement with the National Science Foundation.

\begin{deluxetable}{cccccccc}
\tablecolumns{8}
\tablecaption{Emission lines in J1}
\tablehead{ \colhead{Region} & \colhead{ID} & \colhead{$\lambda_{rest}$} &
\colhead{z} & 
\colhead{$\lambda_{obs}$} & \colhead{FWHM$_{obs}$} & \colhead{FWHM$_{int}$} & \colhead{flux}\\
  (1) & (2) & (3) & (4) & (5) & (6) & (7) & (8)}
\tablecomments{Column (1) -- Regions as defined in
Fig.~\ref{fig:CO}$-$\ref{fig:jmink}.  Column (2) -- Line identification
with wavelength given in Column (3). Column (3) -- Rest-frame wavelengths
in \AA. Column (4) -- Observed wavelengths in $\mu$m. Column (5)
-- Redshift of the line calculated using the wavelength in Column
(3). Column (6) -- Full-width at half-maximum measured in \AA.  Column
(7) -- Intrinsic FWHMs corrected for instrumental resolution in units of
\kms. Column (8) -- Line fluxes in units of 10$^{-20}$ W m$^{-2}$.}
\startdata
J1n &  \ha  & 6563 &  2.5651 $\pm$ 0.00052 &  2.3397 $\pm$ 0.00047 &  23 $\pm$   1 &  259 $\pm$   19 &  2.03$\pm$0.12\\
J1n & [NII] & 6583 &  2.5654 $\pm$ 0.00053 &  2.3471 $\pm$ 0.00048 &  26 $\pm$   2 &  305 $\pm$   32 &  1.05$\pm$0.07\\
J1n & [NII] & 6548 &  2.5658 $\pm$ 0.00058 &  2.3349 $\pm$ 0.00053 &  31 $\pm$   6 &  377 $\pm$   75 &  0.53$\pm$0.04\\
J1n & [SII] & 6717 &  2.5654 $\pm$ 0.00068 &  2.3949 $\pm$ 0.00063 &  26 $\pm$  10 &  296 $\pm$  113 &  0.26$\pm$0.03\\
J1n & [SII] & 6731 &  2.5663 $\pm$ 0.00093 &  2.4001 $\pm$ 0.00087 &  32 $\pm$  18 &  373 $\pm$  209 &  0.29$\pm$0.03\\
J1n & [OIII]& 5007 &  2.5652 $\pm$ 0.00064 &  1.7851 $\pm$ 0.00045 &  14 $\pm$   6 &  188 $\pm$   87 &  0.17$\pm$0.02\\
J1n & \hb   & 4861 &  2.5657 $\pm$ 0.00061 &  1.7333 $\pm$ 0.00041 &  11 $\pm$   5 &  118 $\pm$   58 &  0.13$\pm$0.02\\
J1n & [OII] & 3727 &  2.5656 $\pm$ 0.00085 &  1.3289 $\pm$ 0.00044 &  16 $\pm$   8 &  349 $\pm$  179 &  0.21$\pm$0.03\\
\hline
J1s & \ha   &6563 &  2.5656 $\pm$ 0.00052 &  2.3401 $\pm$ 0.00048 &  22 $\pm$   2 &  240 $\pm$   25 &  1.09$\pm$0.07\\
J1s & [NII] &6583 &  2.5658 $\pm$ 0.00056 &  2.3474 $\pm$ 0.00051 &22 $\pm$   4 &  238 $\pm$   51 &  0.47$\pm$0.04\\
J1s & [NII] &6548 &  2.5667 $\pm$  0.0013 &  2.3354 $\pm$  0.0012 &  66 $\pm$  27 &  835 $\pm$  348 &0.43$\pm$0.03\\
J1s & [SII] &6717 &  2.5726 $\pm$ 0.00085 &  2.3997 $\pm$ 0.00079 &  33 $\pm$  15 &  388 $\pm$  181 &0.28$\pm$0.02\\
J1s & [SII] &6731 &  2.5578 $\pm$ 0.00066 &  2.3944 $\pm$ 0.00062 &  14 $\pm$   9 &  104 $\pm$   68 & 0.14$\pm$0.03\\
J1s & \hb   & 4861 &  2.5658 $\pm$ 0.00060 &  1.7333 $\pm$ 0.00041 &  14 $\pm$   5 &  194 $\pm$   72 & 0.17 $\pm$0.02\\
J1s & [OIII]&5007 &  2.5655 $\pm$ 0.00069 &  1.7853 $\pm$ 0.00048 &  13 $\pm$   7 &  178 $\pm$  102 &  0.11$\pm$0.02\\
J1s & [OII] &3727 &  2.5669 $\pm$ 0.00094 &  1.3294 $\pm$ 0.00049 &23 $\pm$  10 &  506 $\pm$  216 &  0.23$\pm$0.02\\
\hline
J2 & [OIII] & 5007 &  2.5636 $\pm$ 0.00057 &  1.7843 $\pm$ 0.00040 & 13 $\pm$3 & $<$50 &  0.28 $\pm$0.04\\
J2 & [OII]  & 3727 &  2.5649 $\pm$ 0.0007& 1.3286$\pm$0.0003  &16 $\pm$5 & $<$50       & 0.09$\pm$0.007\\
J2 & \hb    & 4861 &  0 &  0 & 0 & 0  &  $<$ 0.09 \\
J2 & \ha    & 6563 &  2.5635 $\pm$ 0.00055 &  2.3387 $\pm$ 0.00050 &14 $\pm$3 & $<$50 &  0.31$\pm$0.04 \\
J2 & [NII]  & 6583 &  2.5628 $\pm$  0.0013 &  2.3454 $\pm$  0.0012 &24$\pm$29& $<$50  &  0.07$\pm$0.02
\enddata
\label{tab:emissionlines}
\end{deluxetable}

\end{document}